\documentclass[reqno,12pt]{article}
\usepackage{amsthm,amsmath,amssymb,graphicx,setspace,verbatim,harvard}
\usepackage[small]{caption}
\usepackage{subcaption}
\usepackage{wrapfig}
\usepackage{multirow}
\usepackage{rotating}
\usepackage{color}
\usepackage{adjustbox}

\newcommand{\bzero}{\mbox{\boldmath $0$}}
\newcommand{\ba}{\mbox{\boldmath $a$}}

\newcommand{\bm}{\mbox{\boldmath $m$}}

\newcommand{\by}{\mbox{\boldmath $y$}}

\newcommand{\bE}{\mbox{\boldmath $E$}}

\newcommand{\bI}{\mbox{\boldmath $I$}}
\newcommand{\bJ}{\mbox{\boldmath $J$}}
\newcommand{\bL}{\mbox{\boldmath $L$}}

\newcommand{\bQ}{\mbox{\boldmath $Q$}}
\newcommand{\bR}{\mbox{\boldmath $R$}}

\newcommand{\bV}{\mbox{\boldmath $V$}}

\newcommand{\bX}{\mbox{\boldmath $X$}}

\newcommand{\cL}{{\cal L}}

\newcommand{\cN}{{\cal N}}

\newcommand{\hZ}{\hat{Z}}

\newcommand{\eps}{\varepsilon}

\newcommand{\hdelta}{\hat{\delta}}

\newcommand{\beps}{\mbox{\boldmath $\varepsilon$}}

\newcommand{\bbeta}{\mbox{\boldmath $\beta$}}
\newcommand{\bsbeta}{\mbox{\scriptsize \boldmath $\beta$}}

\newcommand{\bgamma}{\mbox{\boldmath $\gamma$}}
\newcommand{\bdelta}{\mbox{\boldmath $\delta$}}
\newcommand{\blambda}{\mbox{\boldmath $\lambda$}}
\newcommand{\bmu}{\mbox{\boldmath $\mu$}}

\newcommand{\btau}{\mbox{\boldmath $\tau$}}

\newcommand{\bSigma}{\mbox{\boldmath $\Sigma$}}

\newcommand{\bomega}{\mbox{\boldmath $\omega$}}

\newcommand{\bty}{\mbox{\boldmath $\tilde{y}$}}

\newcommand{\E}{\mbox{E}}
\newcommand{\Var}{\mbox{Var}}

\newcommand{\bdm}{\begin{displaymath}}
\newcommand{\edm}{\end{displaymath}}
\newcommand{\beq}{\begin{equation}}
\newcommand{\eeq}{\end{equation}}

\renewcommand{\th}{^{\mbox{\scriptsize th}}}

\long\def\symbolfootnote[#1]#2{\begingroup%
\def\thefootnote{\fnsymbol{footnote}}\footnote[#1]{#2}\endgroup}

\newcommand{\white}[1]{\textcolor{white}{#1}}

\begin{document}

\pagenumbering{arabic}
\begin{center}
{\singlespacing
\begin{Large}{\bf
Spatiotemporal Modeling of Node Temperatures in Supercomputers\\}
\vspace{.25in}
\end{Large}

Curtis B Storlie$^\dag$, Brian J Reich$^\ddag$, William N Rust$^\dag$, Lawrence O Ticknor$^\dag$, \\
Amanda M Bonnie$^\dag$, Andrew J Montoya$^\dag$, Sarah E Michalak$^\dag$ \\[.15in]

$^\dag$ Los Alamos National Laboratory\\
$^\ddag$ North Carolina State University\\

\begin{abstract}
  Los Alamos National Laboratory (LANL) is home to many large supercomputing clusters.  These clusters require an enormous amount of power ($\sim$500-2000 kW each), and most of this energy is converted into heat.  Thus, cooling the components of the supercomputer becomes a critical and expensive endeavor.  Recently a project was initiated to investigate the effect that changes to the cooling system in a machine room had on three large machines that were housed there.  Coupled with this goal was the aim to develop a general good-practice for characterizing the effect of cooling changes and monitoring machine node temperatures in this and other machine rooms.  This paper focuses on the statistical approach used to quantify the effect that several cooling changes to the room had on the temperatures of the individual nodes of the computers.  The largest cluster in the room has 1,600 nodes that run a variety of jobs during general use.  Since extremes temperatures are important, a Normal distribution plus generalized Pareto distribution for the upper tail is used to model the marginal distribution, along with a Gaussian process copula to account for spatio-temporal dependence.  A Gaussian Markov random field (GMRF) model is used to model the spatial effects on the node temperatures as the cooling changes take place.  This model is then used to assess the condition of the node temperatures after each change to the room.  The analysis approach was used to uncover the cause of a problematic episode of overheating nodes on one of the supercomputing clusters.  This same approach can easily be applied to monitor and investigate cooling systems at other data centers, as well.

\vspace{.075in}

\vspace{.075in}
\noindent
{\em Running title}: Spatiotemporal Modeling of Node Temperatures in Supercomputers

\vspace{.075in}
\noindent
{\em Corresponding Author}: Curtis Storlie, \verb1cbstorlie@gmail.com1

\end{abstract}
}

\end{center}


\vspace{-.35in}
\section{Introduction}
\vspace{-.1in}

\begin{figure}[t!]
\vspace{-.13in}
\centering
\caption{Layout of Mustang, Moonlight, and Pinto in Machine Room 341.}
\vspace{-.15in}
\includegraphics[width=.64\textwidth,height=.44\textheight]{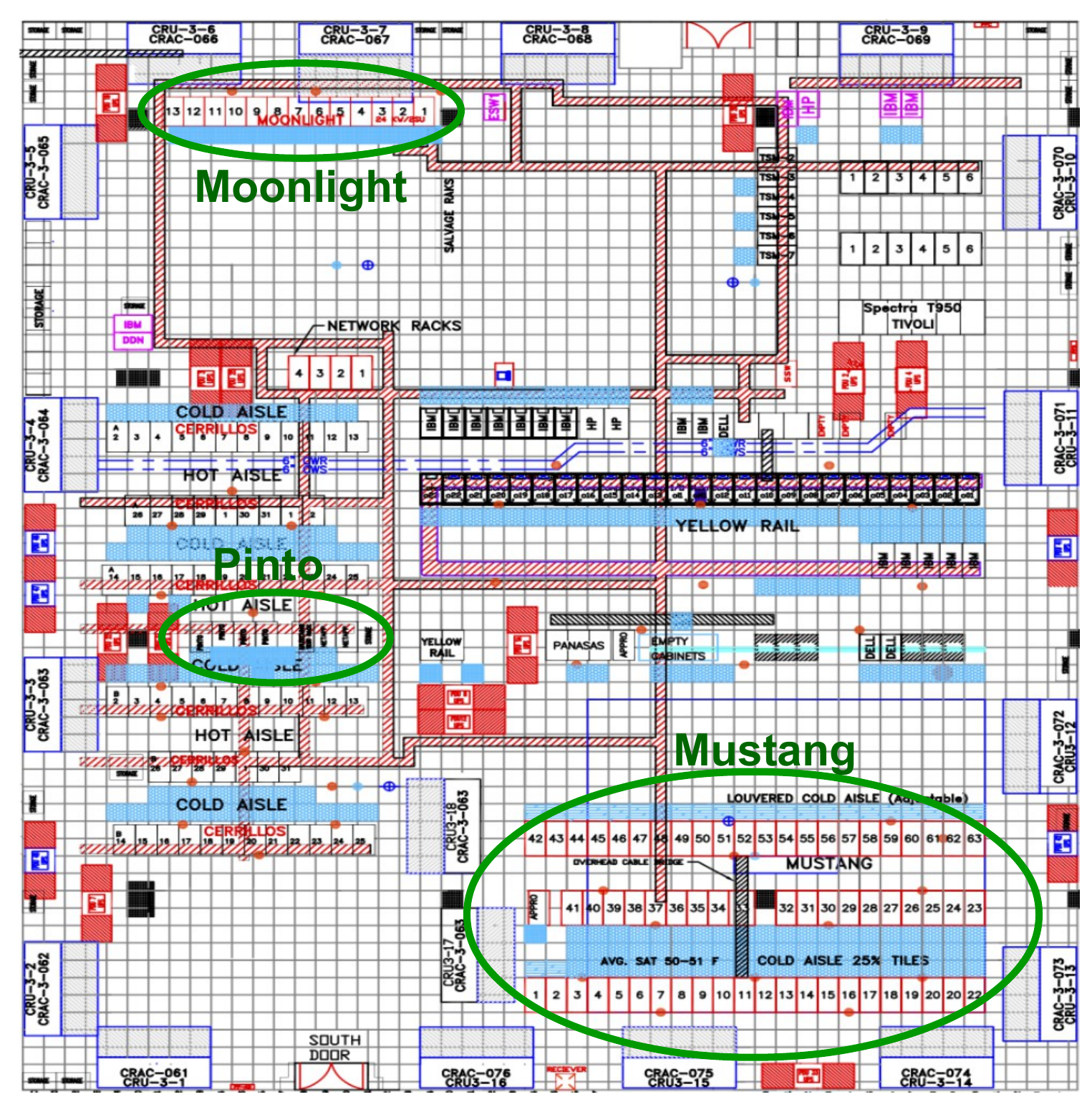}
\label{fig:341_layout}
\vspace{-.175in}
\end{figure}

The cooling of components in high performance computing (HPC) centers is a critical issue.  Most of the hundreds of kilowatts of energy used to power a large supercomputing machine are converted into heat.  This heat must be taken away from the components in order to prevent overheating and damage.  Thus, cooling strategy is a major consideration for a large data center.  Los Alamos National Laboratory (LANL) is home to many large supercomputing machines.  Several machines are generally housed together in a single machine room.  This paper focuses on machine room 341, which was (as of January 2014) home to the three computing machines listed in Table~\ref{tab:machine_specs}.  In room 341, cool air is pumped into the machine room through perforated tiles in the floor.  The cool air is then sucked into the machine components (e.g., compute nodes) and hot air is blown out.  The layout of room 341 with these machines circled is provided in Figure~\ref{fig:341_layout}.  A project was initiated with the goal of investigating the effect that cooling changes had on the machines, while ensuring that the components of these machines are not subject to overheating.  A further goal was to develop a general good-practice procedure for investigating the effect of cooling changes in other machine rooms and for monitoring room 341 as layout changes occur (e.g., the installation of a new cluster).  This paper focuses primarily on the statistical challenges involved in accomplishing this goal.

\renewcommand{\tabcolsep}{.05in}
\begin{table}[t!]
\vspace{-.13in}
\centering
\caption{Temperature thresholds and specs for the three compute machines in room 341.}
\vspace{.0in}
\label{tab:machine_specs}
\vspace{-.1in}
{\small
\begin{tabular}{|l|c|c|c|r|c|r|}
\hline
   & \multicolumn{3}{c|}{Thresholds} &\multicolumn{3}{c|}{Machine Specs} \\
\cline{2-7} 
Cluster &  Warning & High & Critical & \# Nodes &  Cores$/$Node & Load (kW) \\
\hline
Mustang   & 59$^\circ\:$C & 65$^\circ\:$C & $\;\:$70$^\circ\:$C &  1,600$\;\;$ & 24 & 730$\;\;$ \\
Moonlight & 89$^\circ\:$C & 95$^\circ\:$C & 100$^\circ\:$C &  614$\;\;$ & 16 & 530$\;\;$ \\
Pinto     & 89$^\circ\:$C & 95$^\circ\:$C & 100$^\circ\:$C &  162$\;\;$ & 16 & 67$\;\;$ \\
\hline
\end{tabular}
}
\vspace{-.15in}
\end{table}

Room 341 has 18 computer room air conditioning (CRAC) units, 16 along the sides of the room and two more directly left of Mustang in Figure~\ref{fig:341_layout}.  All of the 18 CRAC units were operating at the start of the project.  The hypothesis from the facilities team was that several of the CRAC units could be shut off while still providing the pressure necessary to allow adequate airflow to the supercomputer nodes.  Also, the current cooling supply temperature of 12.8$^\circ\:$C was believed to be too conservative, and that this could be increased without having much impact on the cooling of the compute machine components.
Machine room 341 has other components (such as data storage, networking, etc.) in addition to the compute machines, but the compute machines are the biggest heat producers and the primary concern.  For brevity, we focus our attention in this paper on the largest of the three machines, the Mustang cluster.  The Mustang cluster is the most complicated from a statistical modeling perspective, and it is also the most problematic of the three machines, from a heat perspective.  The other two machines have also been modeled with the same approach to be described here.

\begin{figure}[t!]
\vspace{-.1in}
  \centering
\caption{Rack and node layout of the Mustang cluster.}
  \vspace{-.1in}
\label{fig:Mustang_layout}
\includegraphics[width=.9\textwidth]{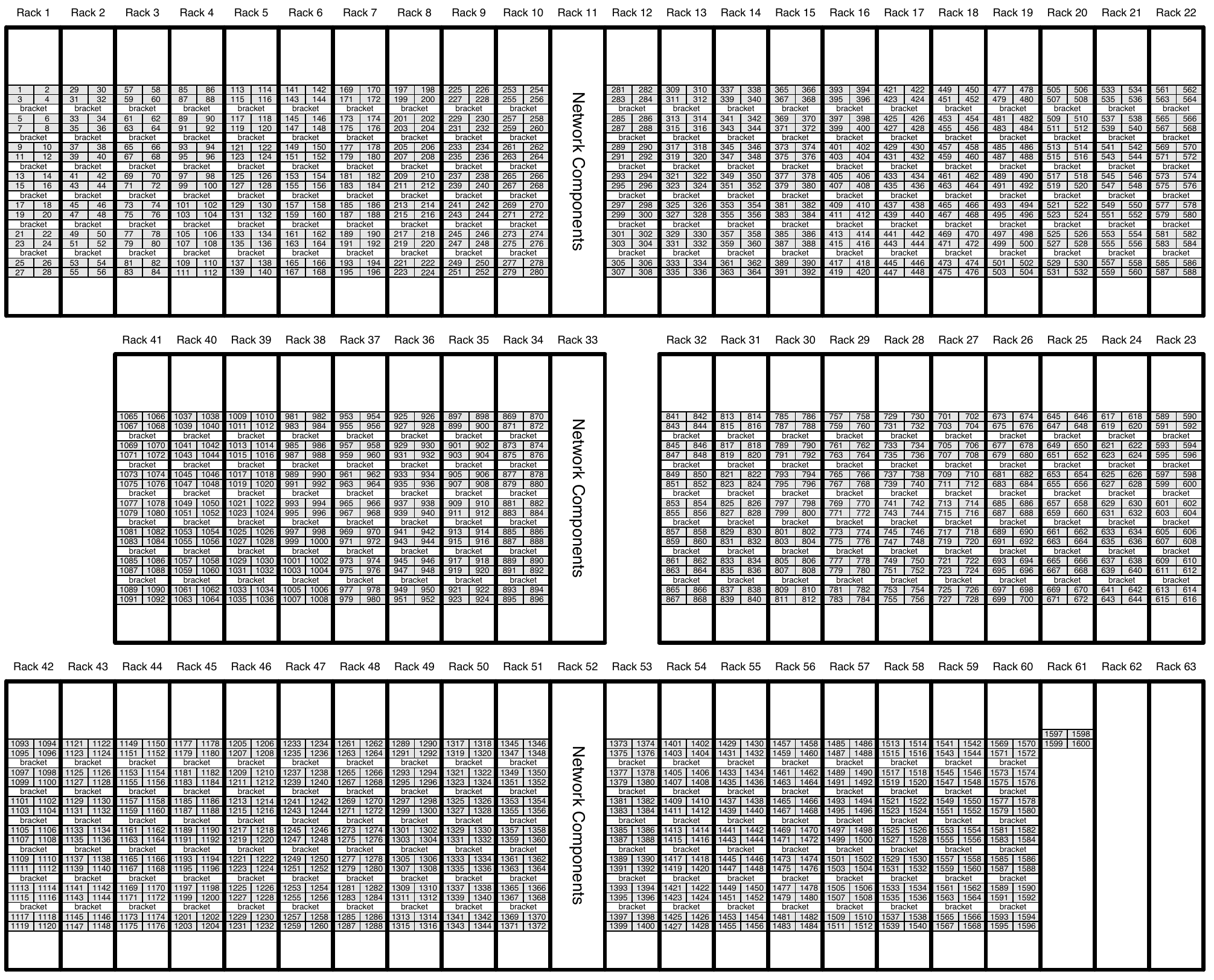}
\vspace{-.1in}
\end{figure}

As described in Table~\ref{tab:machine_specs}, the Mustang machine has 1,600 compute {\em nodes}, spread out among 58 compute racks each housing 28 nodes (with the exception of the last compute rack which has only four nodes).  There are actually a total of 63 racks, but only 58 of them house compute nodes (the others are empty or contain network and file system components).  The racks are laid out as three rows as can be seen in Figure~\ref{fig:341_layout} and further in Figure~\ref{fig:Mustang_layout}, which shows the layout of the nodes within each of the racks.  

Each node is aware of its temperature, measured at the CPU, and it will report a warning if it exceeds 59$^\circ\:$C.  A node will report a high temperature warning if it exceeds 65$^\circ\:$C, but allow the current job to finish.  Once the current job is finished, that node would then be removed from the available nodes in the job queue and inspected for any hardware issues that may have led to a high temp.  If a node reaches 70$^\circ\:$C, the current job is killed immediately and the node is removed for inspection.  Under normal operation it is undesirable for nodes to reach 65$^\circ\:$C very often, or to reach 70$^\circ\:$C at all.

The plan used to investigate the effect on the machine(s) due to changes to the cooling system was to (i) develop a statistical model for node temperatures over time and space as a function of cooling supply temperature and other ``effects'' to the room, e.g., turning off CRAC units or the installation of a new cluster, and (ii) use the model to assess the current {\em state-of-the-machine} and assess the feasibility of another cooling change.  For this purpose, the current state-of-the-machine was defined to be a 95\% credible bound for the maximum temp that would be achieved if launching a new HPL job on the entire machine and letting it run for a full day.  The HPL job is a compute intensive program that performs an LU decomposition of a large matrix and uses it to solve a linear system of equations \cite{Dongarra11}.  HPL tests were conducted during designated service times (DSTs) so as not to interfere with user jobs.  DSTs occur roughly once a month, but tests occurred more on the order of every two or three months as it proved more difficult than anticipated to obtain time for experimentation during the DSTs.
During the designated HPL tests, which typically ran for about two hours, node temperatures were collected every minute.
HPL is somewhat synthetic in that it is not truly representative of a real {\em production} job (i.e., a real scientific computing job) that might run on Mustang, but it is representative of a worst-case computation that a node might experience as part of a regular user's production job.  The somewhat conservative benchmark program, however, removes the user/job variability and allows for a much easier identification of the changes that nodes may experience due to effects of interest.

\begin{figure}[t!]
  \vspace{-.1in}
  \centering
  \caption{Temperature data from six selected Mustang nodes during the baseline experiment.}
\vspace{-.2in}
\includegraphics[width=.7\textwidth]{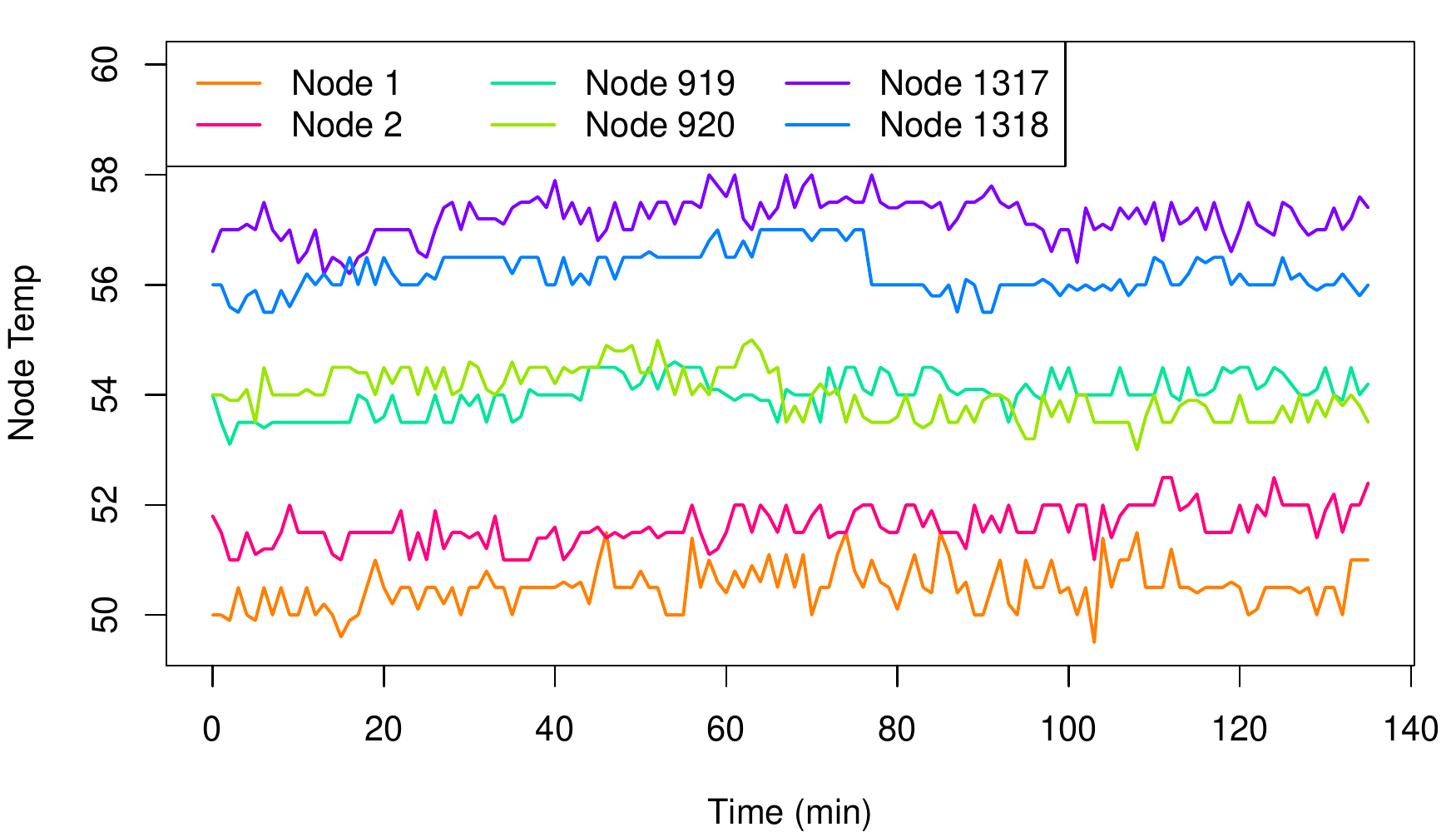}
\label{fig:baseline_data}
\vspace{-.1in}
\end{figure}

Some of the temperature data from the HPL experiment to establish a {\em baseline} for Mustang are displayed in Figure~\ref{fig:baseline_data}.  HPL was run on all 1,600 nodes, but for ease of display, only temperatures from six selected nodes $(1,2,919,920,1317,1318)$ during the experiment are plotted here.  The temperature time series for neighboring nodes cluster together which is not a fluke as there is substantial spatial correlation between node temperatures in these data.  A video of the node temperatures during the course of the baseline experiment along with complete temperature data for all HPL experiments used in this paper are available at the journal website.  The time points (during an experiment) are approximately 1 minute apart, ranging from about 50 seconds to 70 seconds due to the timing of the query from the server to the 1,600 nodes and how busy the nodes are at that time, etc.  Also, the message to the nodes can be lost and the node may not report its temperature for that minute; this happened approximately 10\% of the time. The various experiments to be analyzed in Section~\ref{sec:analysis} were conducted several weeks apart from each other, so there are large time gaps in the data as well.

The node temperature data display both spatial and temporal variation.   In general, the space and time domains can be either continuous or discrete which leads to four different settings \cite{Cressie2011}.  Much of the methodological development for spatiotemporal data has considered the discrete time domain (e.g., \citename{Waller1997} \citeyear*{Waller1997} for discrete space; and \citename{Handcock1994} \citeyear*{Handcock1994}; \citename{Gelfand2005} \citeyear*{Gelfand2005} for continuous space).   Elaborate correlation models have also been proposed for continuous space with continuous time \citeaffixed{Cressie1999,Gneiting2002}{e.g.,}.  Since the nodes constitute a discrete spatial domain and temperature changes continuously with time, the setting in this paper corresponds to a discrete space with continuous time. This framework is less common, but has been explored by \citeasnoun{MacNab2007} and \citeasnoun{Quick2013}.   The current application and corresponding methodology used here are different in that they place an emphasis on extreme values.

As will be seen in Section~\ref{sec:analysis}, the node temperature distribution has a very heavy upper tail when running HPL.  Being able to represent these extremes well statistically is critical to the characterization of the state-of-the-machine, thus, we employ methods from spatial extreme value analysis \cite{Davison12}.  Standard extreme value approaches isolate only extremes, either those above a predetermined threshold or the maximum of a group of observations \cite{Coles05}.  In contrast, our analysis is facilitated by a model for all observations, both extreme and non-extreme.  The marginal distribution of node temperature is asumed to be a combination of Gaussian for non-extremes and a generalized Pareto distribution (GPD) for the tail \cite{Frigessi02,Carreau09a,Carreau09b,Reich13}.  The parameters are allowed to vary spatially and by experimental conditions, providing a means to assess the effect on cooling due to different scenarios and help inform cooling strategy.  The structure of the node layout is leveraged to develop a Markov random field model \cite{Rue01,Rue05,Li09} for the spatial effects to maintain computational feasibility for large machines.  Dependence in the residual process is also considered.  A natural model for extremal spatial and/or temporal dependence is the max-stable process model \citeaffixed{Smith90,Kabluchko09,Padoan10,Wadsworth12,Reich12b,Huser13,Huser14,Wadsworth14}{e.g.,}.  However, max-stable processes are motivated primarily for analysis of extremes alone and computation is tedious for the large datasets we consider here.  Instead, we use a Gaussian process (GP) copula (see, e.g., \citeasnoun{Nelson99}, for a general review and \citeasnoun{Sang09}, for an application to spatial extremes) which is computationally efficient and demonstrate that this approach is sufficient to capture the important features of our data.

Sophisticated statistical modeling has been used recently to address issues of power consumption \cite{Storlie14power} and reliability \cite{Storlie12a,Michalak12} of HPC systems.  However, this is the first attempt to model spatiotemporal node temperatures in supercomputers.  This model is then used here to identify causes of temperature issues and assess various cooling strategies.
The rest of the paper is laid out as follows.  Section~\ref{sec:node_temp_model} describes the hierarchical Bayesian model for the node temperatures.  Section~\ref{sec:analysis} provides an in depth analysis of the effect that room changes have on the Mustang nodes, and Section~\ref{sec:conclusions} concludes the paper.  This paper also has online supplementary material containing data and Markov chain Monte Carlo (MCMC) details.


\vspace{-.2in}
\section{Statistical Model for Node Temperatures}
\vspace{-.1in}
\label{sec:node_temp_model}


\vspace{-.15in}
\subsection{Model Description}
\vspace{-.1in}
\label{sec:model_descr}

The proposed model allows the mean node temp to change (e.g., due to supply temperature and/or other room/node change covariates $x_j$) according to spatial random effects $\beta_j$.  There is also a residual process $\delta$ to capture the remaining variation in space and time. A thorough assessment of the feasibility of assumptions made in the model below for application to the node temperatures of Mustang can be found in Section~\ref{sec:assumptions}.  Specifically, the temperature of node $s$ at time $t$ is
\vspace{-.15in}\beq
y(s,t) = \beta_0(s) + \sum_{j=1}^J \beta_j(s) x_j(s,t) + \delta(s,t) + \eps_{s,t}, 
\label{eq:node_model}
\vspace{-.15in}\eeq
for $s \in \{1,\dots,S\}$ and $t \in [0, \infty)$, where $\bbeta_j = [\beta_j(1), \dots, \beta_j(S)]' \sim N(\mu_j\bJ_{\!S},\bSigma_j)$, with $\bJ_{\!S}$ a vector of all ones of length $S$ (the number of nodes), and  $\eps_{s,t} \stackrel{iid}{\sim} N(0,\sigma^2)$ represents measurement error.
The model for the residual process $\delta(s,t)$ requires some care.  It must accurately represent the extremes of the distribution since they will have a large influence on the state-of-the-machine.  Thus, $\delta(s,t)$ is assumed to be a dependent process with a marginal distribution that can flexibly account for extremes as described below.  For computational convenience, it will be assumed that the $\delta_s = \delta(s,\cdot)$ process is independent of $\delta_{s'}$ for $s \neq s'$. Upon examination of the residuals, this assumption is entirely reasonable for this application after accounting for the spatially varying effects, $\beta_j$s; see Section~\ref{sec:assumptions}.  Thus, we present a model for time dependent $\delta_s$, independent across $s$, below.  If there were significant spatial dependency in $\delta$, one could consider using product correlation, which has nice computational advantages, but some theoretical drawbacks \cite{Stein05}, or use any multivariate method such as a factor or principle components model.

The {\em tail} of a wide class of distributions is well approximated by a GPD \cite{Coles05}.  Hence, we let the marginal density of $\delta_s$, $f_\delta$, be the density of a normal distribution that switches to a GPD density for the upper tail, i.e., 
\vspace{-.18in}\beq
f_\delta(y) = \phi\left(\frac{y}{\upsilon}\right) I_{\{y \leq \kappa \upsilon\}} + 
\left[1-\Phi(\kappa \upsilon) \right]g(y-\kappa \upsilon; \xi, \eta) I_{\{y > \kappa \upsilon\}},
\label{eq:F_delta}
\vspace{-.18in}\eeq
where (i) $\phi$ and $\Phi$ are the standard normal density and CDF, respectively, and $g(\cdot;\xi,\eta)$ is the GPD density with shape parameter $\xi$ and scale parameter $\eta$ (and threshold parameter equal 0). Specifically,
\bdm
g(x;\xi,\eta) = \frac{1}{\eta} \left( 1 + \frac{\xi x}{\eta} \right)^{-\frac{1}{\xi} - 1}.
\edm
Thus $f_\delta$ is assumed normal mean 0, variance $\upsilon^2$, until a point, $\kappa$ standard deviation units above the mean, at which point the GPD takes over.  The parameter $\eta$ can be chosen as a function of $\kappa$ to enforce a continuity of $f_\delta$ at $\kappa \upsilon$ which is done here.  That is, we assume
\vspace{-.20in}\bdm
\eta = \frac{1-\Phi(\kappa)}{\phi(\kappa)}
\vspace{-.16in}\edm
so that $f_\delta(\kappa \upsilon) = \lim_{x \downarrow \kappa \upsilon} f_\delta(x)$

A model is now described that creates a $\delta_s$ {\em process} with marginal distribution $f_\delta$ in (\ref{eq:F_delta}).  A common approach to create a dependent process $\delta_s$ that has a desired marginal distribution is to make use of a copula as in \citeasnoun{Sang09} and \citeasnoun{Reich12}.  A copula is a dependent process with uniform marginals, thus an inverse CDF transform of a copula will then produce the desired marginals.  For example, assume that
\vspace{-.18in}\beq
\delta_s(t) = F_\delta^{-1} \left[\Phi\left( Z_s(t) \right)\right], \; s=1,\dots,S,
\label{eq:meta_delta}
\vspace{-.2in}\eeq
where $F_\delta(x) = \int_{-\infty}^xf_\delta(x)$ is the desired marginal CDF, and $Z_s$ is a stationary Gaussian process (GP) with a mean 0, variance 1.  For any point $t$, $Z_s(t)$ has a standard normal distribution, and $U_s(t)=\Phi(Z_s(t))$ necessarily has a Unif$(0,1)$ distribution, i.e., the inverse-CDF transform yields the desired marginal distribution for $\delta_s(t)$.  The process $U_s(t)$ is a Gaussian-copula and the resulting  $\delta_s$ is a meta-GP \cite{Demarta05}.  Here we assume that the correlation model for $Z_s$ is exponential, i.e.,
\vspace{-.22in}\beq
K_{\delta}\left(t,t'\right) =  \exp\{-\theta |t-t'|\}.
\label{eq:delta_cov}
\vspace{-.22in}\eeq

The meta-GP in (\ref{eq:meta_delta}) is a dependent process with flexible tail behavior in the marginals, but it is well known that a meta-GP fails to allow for asymptotic dependence in extreme values \cite{Frahm05}, i.e., $\chi(c) = \Pr(Z_s(t) > c \mid Z_s(t-1) > c) \rightarrow 0$ as $c \rightarrow \infty$ for bivariate Gaussian random variables with correlation less than one \cite{Coles13}.  \citeasnoun{Demarta05} recommend the use of a $t$-copula to allow for such tail dependence.  However, the Mustang temperature data provided little evidence of asymptotic tail dependence; a detailed investigation is provided in Section~\ref{sec:assumptions}.  Thus, a Gaussian copula was deemed sufficient for this analysis.

\begin{figure}[t!]
\vspace{-.2in}
  \centering
\caption{Node neighbor relationships for Mustang}
\vspace{-.1in}
\label{fig:node_topology}
\includegraphics[width=.665\textwidth]{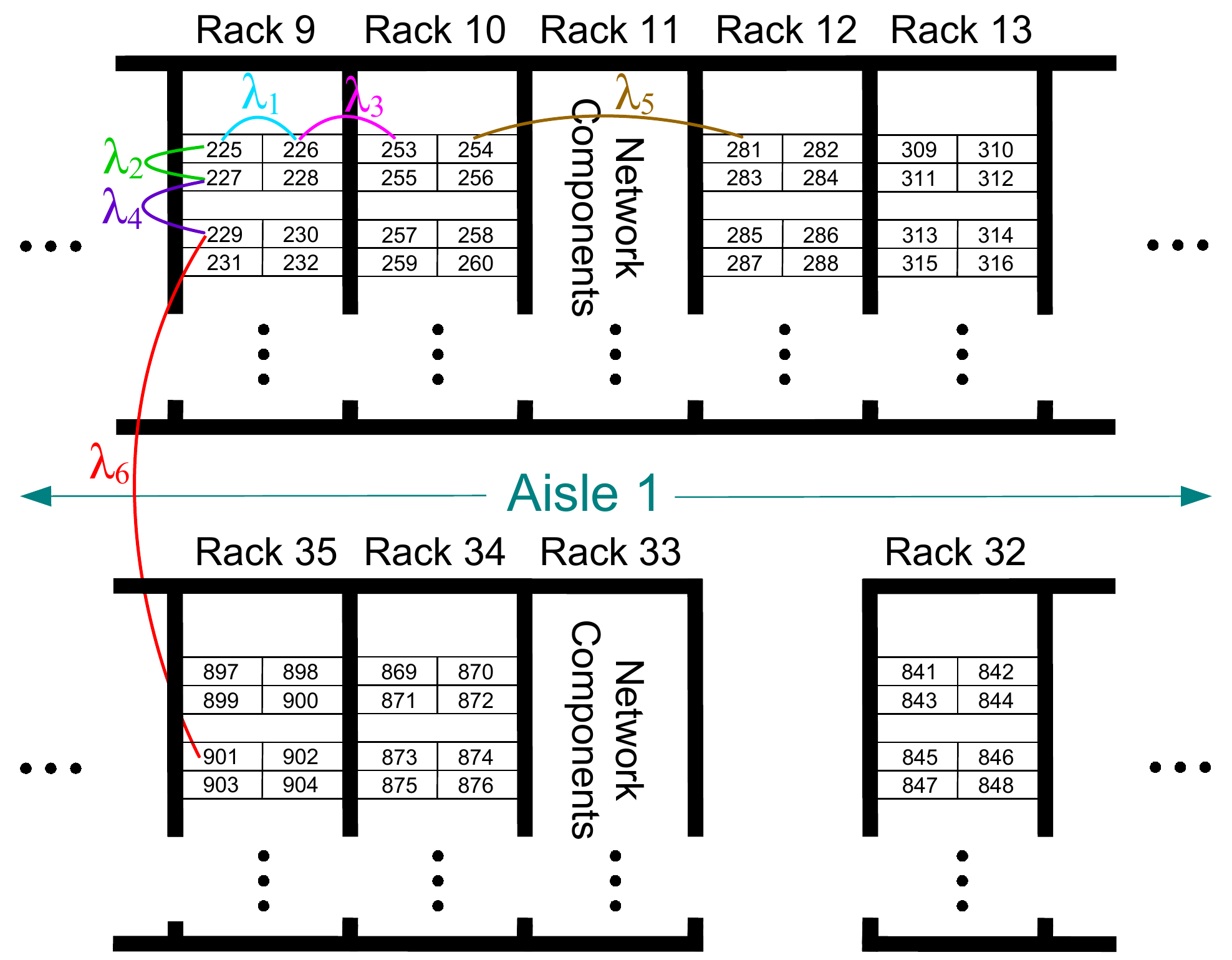}
\vspace{-.1in}
\end{figure}

The model in (\ref{eq:node_model}) has several $\bbeta_j$ in space where the spatial dimension (for Mustang) is 1,600 and can be more than 10,000 nodes for the largest machines at LANL.  Since a traditional multivariate normal model requires ${\cal O}(N^3)$ operations for likelihood evaluation and/or realizations, we assume a Gaussian Markov random Field (GMRF) model to alleviate computational burden.
Each process $\bbeta_j$ is assumed to be a GMRF with conditionally autoregressive (CAR) representation
\vspace{-.18in}\beq
\E(\beta_{j,s} \mid \bbeta_{j,-s}) = \mu_j + \varphi \; \frac{\sum_{l=1}^L \sum_{r \in \cN_{s,l}} \lambda_l \left( \beta_{j,r} - \mu_j \right)}{\sum_{l=1}^L \lambda_l n_{s,l}}
\label{eq:cond_mean}
\vspace{-.2in}\eeq
and precision
\vspace{-.05in}\bdm
\left[\Var(\beta_{j,s} \mid \bbeta_{j,-s}) \right]^{-1} = \frac{\tau_j}{\varphi} \sum_{l=1}^L \lambda_l n_{s,l},
\vspace{-.05in}\edm
where $\cN_{s,l}$ is the set of neighbors of type $l$ for node $s$ and $n_{s,l} = |\cN_{s,l}|$ is the number of neighbors of type $l=1,\dots,L$ for node $s$.  The neighborhood relationships for Mustang (with $L=7$) are illustrated in Figure~\ref{fig:node_topology}, which is a zoomed in version of Figure~\ref{fig:Mustang_layout}. Horizontally neighboring nodes within the same rack are neighbors of type 1 and are given an autoregression coefficient of $\lambda_1$, while vertically neighboring nodes (which are not separated by a shelf) are neighbors of type 2.  If nodes are horizontal neighbors, but with a rack boundary in between them, they are type 3 neighbors.  Likewise, if nodes are vertical neighbors, but with a shelf in between, they are type 4 neighbors. If nodes have a rack of network components in between them, but are otherwise horizontally aligned, they are type 5 neighbors.  If nodes have the same geography within rows 1 and 2 but are directly across aisle 1 from one another, then they are type 6 neighbors.  Finally, because of the different orientation of the front/back of the nodes and thus cooling in each aisle, neighbors between rows 2 and row 3 (across aisle 2) are treated as type 7 neighbors.

The conditional mean of $\beta_{j,s}$ is $\varphi$ times a weighted {\em average} of the neighbors, where the $\lambda_l$ control the relative weights of the $l\th$ neighbor type in the average.  For identifiability, it is assumed that $\lambda=[\lambda_1,\dots,\lambda_L]' \sim \mbox{Dirichlet}(\ba_\lambda)$ so that the $\lambda_l$ sum to 1.  The parameter $\varphi \sim \mbox{Beta}(A_\varphi, B_\varphi)$ acts as a single autoregression coefficient on the weighted average of the neighbors.  The $\lambda_l$ can also be compared to determine the relative importance of the neighbor types in the dependency.  The precision is proportional to the number of neighbors (and their weights) that go into the weighted average (e.g., nodes on the boundary will have a larger conditional variance because they have fewer neighbors) and $\tau_j$ scales the overall precision of $\beta_j$.  The CAR representation in (\ref{eq:cond_mean}) results in the following precision matrix $\bQ_j$ for $\bbeta_j$,
\vspace{-.2in}\beq
(\bQ_j)_{r,s} = \left\{
\begin{array}{ll}
  - \tau_j \lambda_l   &  \mbox{if $r \in \cN_{s,l}$} \\
  \frac{\tau_j}{\varphi} \sum_{l=1}^L \lambda_l n_{s,l}  &  \mbox{if $r = s$} \\
  0 & \mbox{otherwise.}\\
\end{array}
\right. 
\label{eq:Q}
\vspace{-.15in}\eeq
In order to assure a positive definite (PD) $\bQ_j$ it is common to assume diagonal dominance (DD), i.e., $\sum_{r\neq s} (\bQ_j)_{r,s} < (\bQ_j)_{s,s}$, a sufficient (but not necessary) condition for PD.  It can be seen in (\ref{eq:Q}) that $\bQ_j$ is DD if and only if $\varphi<1$.

Perhaps a more common formulation \citeaffixed{Reich07}{e.g.,} for a CAR model with multiple neighbor types than that in (\ref{eq:cond_mean}) would be
\vspace{-.2in}\begin{eqnarray}
\E(\beta_{j,s} \mid \bbeta_{j,-s}) & \!= \!&\mu_j +  \frac{\sum_{l=1}^L \sum_{r \in \cN_{s,l}} \rho_l \left( \beta_{j,r} - \mu_j \right)}{\sum_{l=1}^L n_{s,l}}, \label{eq:cond_mean2} \\
\left[\Var(\beta_{j,s} \mid \bbeta_{j,-s}) \right]^{-1} & \!= \!& \tau_j \sum_{l=1}^L n_{s,l}. \nonumber  \\[-.475in] \nonumber
\end{eqnarray}
DD in the case of (\ref{eq:cond_mean2}) requires $\sum_l n_{s,l} \rho_l < \sum_l n_{s,l}$, for all $s$, which is cumbersome to impose exactly.  Thus, it is commonly assumed that all $\rho_l < 1$, which satisfies the DD condition, but this is much more restrictive than DD.  Hence, the formulation in (\ref{eq:cond_mean}) is more flexible and can allow more weight on some neighbor types than is possible in (\ref{eq:cond_mean2}).
It is well-documented in the literature, that the $\rho_l$ parameters need to have values {\em very} close to 1 to capture any reasonable spatial association between neighbors.
Indeed, when the model in (\ref{eq:cond_mean2}) was applied to the node temperature data, all of the $\rho_l$ were very close to 1 and the resulting {\em correlations} between neighboring nodes of all neighbor types were near $\sim 0.6$.  Meanwhile, the $\beta_j$ in the posterior had much higher (empirical) correlation ($\sim0.9$) for neighbor types 1-4, indicating a poor fit for the GMRF in (\ref{eq:cond_mean2}).

The issue with the formulation in (\ref{eq:cond_mean2}) is that in order to get high $(>0.8)$ correlation between nodes of a given neighbor type $l'$, it is not good enough to have $\rho_{l'}$ near 1, rather {\em all} of the $\rho_l$ must be very near 1.  However, this implies high correlation in all directions.  What is missing is the ability to allow some of the $\rho_l >1$ which is essentially what the formulation in (\ref{eq:cond_mean}) is allowing for in a convenient parameterization.  
In contrast, the model in (\ref{eq:cond_mean}) fit to the node temperature data results in a $\varphi$ very close to 1, but with distinct separation in the $\lambda_l$, and much higher correlation ($\sim0.9$) between neighbor types 1-4.  Thus, the model in (\ref{eq:cond_mean}) allows for a more flexible fit and better assessment of the relative dependency between neighbor types as seen in Section~\ref{sec:baseline}.

In this problem $\bQ_j$ is a 1,600$\times$1,600 matrix that is sparse with only 9,506 nonzero entries.  It can be a large computational advantage to work
with a the sparse precision matrix resulting from a GMRF for likelihood evaluation or conjugate updates of the $\beta_j$ vector (see the Supplementary Material).  In this case (after permutation) $\bQ_j$ is a banded matrix with a bandwidth of 43 and Cholesky decomposition takes $\sim$0.0006 seconds in R (using the \verb1Matrix1 package) as opposed to $\sim$0.7 seconds (over 1,000 times longer) for a dense matrix.  This decomposition would otherwise be a computational bottleneck as it is needed several times during each MCMC iteration.  
This computational savings will be even more critical when extending this approach to other LANL machines (e.g., the Trinity machine will have more than 10,000 nodes).  



A summary of the model structure and the assumed prior distributions are provided in Table~\ref{tab:priors}.  Relatively diffuse priors were used for all parameters, guided by some intuition and input from the operations team.  For example, the prior for $\mu_j$, the mean of $\beta_j$, implies spatial effects are expected to be smaller in magnitude than $10^\circ\:$C on average.  The prior for $\varphi$ allows for anything in $(0,1)$, but favors larger values.
The prior for $\theta$ essentially allows anything from correlation $\sim 0.99$ to negligible correlation for observations 1 minute apart (on the same node).  The prior for $\kappa$ has a mode of 2 (standard deviation units) and a $99\th$ percentile of 5.8, while the prior for $\xi$ is restricted in this case to be positive (i.e., heavy tail) with a mode of 0.5 and a $99\th$ percentile of 3.  The prior for $\sigma^2$ was constructed based on the expert judgment that the error in temperature measurements should have a standard deviation of about $0.5^\circ\:$C.  Several adjustments (within reasonable ranges) were made to the prior distributions to assess sensitivity.  No significant sensitivity to prior specification was found insofar as its effect on the posterior of $\beta_j$, $\theta$, $\kappa$, or $\xi$.

\begin{table}[t]
\vspace{-.13in}
\begin{center}
\caption{Summary of hierarchical model for node temperature model defined in (\ref{eq:node_model}), (\ref{eq:delta_cov}), and (\ref{eq:Q}), and the specification of prior distributions.}
\label{tab:priors}
\vspace{-.1in}
{\small
  \renewcommand{\tabcolsep}{4pt}
\begin{tabular}{|l|c|c|c|}
  \hline
  Description & Model & Prior Distributions & Specification \\
  \hline
  \hline
  \multirow{8}{*}{Spatial Effects} &  & \multirow{2}{*}{$\mu_j \stackrel{ind}{\sim}  N(M_j, S^2_j)$, $j=0,...,J$} & $M_j= 0$ \\
  &  & & $S^2_j= 100$ \\
  \cline{3-4}
  & \multirow{2}{*}{$\bbeta_j \stackrel{ind}{\sim} N(\mu_j\bJ_{\!\!S}, \bQ_j^{-1})$}& \multirow{2}{*}{$\tau_j \stackrel{iid}{\sim} \Gamma(A_\tau, B_\tau)$, $j=0,...,J$} & $A_\tau= 1$ \\
  & & & $B_\tau=0.5$ \\
  \cline{3-4}
  &  \multirow{1}{*}{as in (\ref{eq:node_model}) and (\ref{eq:Q})} & \multirow{2}{*}{$\lambda_{l} \sim \mbox{Dir}(\ba_\lambda)$} & \multirow{2}{*}{$\ba_\lambda = [1,...,1]'$} \\
  & & & \\
  \cline{3-4}
  & & \multirow{2}{*}{$\varphi \sim \mbox{Beta}(A_\varphi, B_\varphi)$} & $A_\varphi= 5$ \\
  & &  & $B_\varphi= 1$\\
  \hline
  \hline
  & & \multirow{2}{*}{$\upsilon^2 \sim \mbox{IG}(A_\upsilon, B_\upsilon)$} & $A_\upsilon= 5$ \\
  & & & $B_\upsilon=2$ \\
  \cline{3-4}
  & \multirow{2}{*}{$\delta_s \stackrel{iid}{\sim} \mbox{meta-GP}$} & \multirow{2}{*}{$\theta \sim \Gamma(A_\theta, B_\theta)$} & $A_\theta=2$ \\
   \multirow{2}{*}{Residual Process} & & & $B_\theta=2$ \\
  \cline{3-4}
  & \multirow{1}{*}{as in (\ref{eq:F_delta}) and (\ref{eq:meta_delta})} & \multirow{2}{*}{$\kappa \sim \Gamma(A_\kappa, B_\kappa)$} & $A_\kappa=4$ \\
  & & & $B_\kappa=2$ \\
  \cline{3-4}
  & & \multirow{2}{*}{$\xi \sim \Gamma(A_\xi, B_\xi)$} & $A_\xi=2$ \\
  & & & $B_\xi=2$ \\

\hline
\hline
  \multirow{2}{*}{Measurement Error} &  \multirow{2}{*}{$\eps_{s,t} \stackrel{iid}{\sim}N(0, \sigma^2)$} & \multirow{2}{*}{$\sigma^2 \sim \mbox{IG}(A_\sigma, B_\sigma)$} & $A_\sigma= 10$ \\
  & & & $B_\sigma=2$ \\
\hline
\end{tabular}
}
\end{center}
\vspace{-.23in}
\end{table}

\vspace{-.21in}
\subsection{MCMC Algorithm}
\vspace{-.12in}
\label{sec:param_estimation}

The complete list of parameters in the model described in (\ref{eq:node_model}), (\ref{eq:delta_cov}), and (\ref{eq:Q}) is,  
\vspace{-.27in}\beq
\Theta = \left\{
\bbeta,
\bdelta,
\bmu,
\btau,
\blambda,
\upsilon^2,
\theta,
\kappa,
\xi,
\sigma^2
\right\},
\label{eq:param_list}
\vspace{-.28in}\eeq
where $\bbeta = \left\{\beta_j(s), j=1,...,J,\;s=1,...,S \right\}$, $\bdelta=\left\{\delta_s(t_{s,n}), s=1,...,S,\;n=1,...,N_s \right\}$, and $\btau = [\tau_1, \dots, \tau_J]'$.
The posterior distribution of these parameters is approximated via Markov chain Monte Carlo (MCMC).
The complete details of the MCMC algorithm, including full conditional distributions, etc., are provided in the Supplementary Material.  However, an overview is provided here to illustrate the main idea.

The MCMC routine is a typical hybrid Gibbs, Metropolis Hastings (MH) sampling scheme \citeaffixed{Gelman2014}{see e.g.,}.  Conjugate updates are available for all parameters in (\ref{eq:param_list}) with the exception of $\bdelta$, $\blambda$, and $\theta$, which require MH updates.  The $\blambda$ vector was updated via a random walk proposal using a Dirichlet distribution.  The time correlation parameter $\theta$ was updated using a Gaussian random walk on log scale.  If the time-varying residuals $\bdelta$ were assumed to be a GP, then they would have the typical conjugate Gaussian update.  Of course, with the model in (\ref{eq:meta_delta}), $\bdelta$ is no longer a conjugate update, nor can it be integrated out analytically.  However, a GP conditional on ``extreme'' data has no trouble acting extreme.  A GP would just not produce such extreme data, unconditionally, which would result in unrealistic realizations of future temperatures, and thus the reason for the model in (\ref{eq:meta_delta}).  A simple, yet effective approach to the MCMC computation is then to use a conjugate Gaussian update (assuming $\delta_s \sim GP(0,\upsilon^2K_\delta)$) to form a proposal for $\delta_s$ in a MH update.  This approach provided good mixing for the analyses in Section~\ref{sec:analysis} with the benefit of no tuning.  With the efficient GMRF representation for $\beta_j$ in place, the computational bottleneck becomes the updating of $\delta_s$.  However, the independence assumption over space for $\delta_s$ easily allows for parallel updates of each $\delta_s$.
On the largest data set analyzed here ($\sim 700,000$ observations) the MCMC algorithm took $\sim$8 hours for 20,000 iterations (1.5 seconds/iteration) on a 24 core machine with 2.4GHz processors.



\vspace{-.2in}
\section{Analysis of the Mustang Cluster}
\vspace{-.1in}
\label{sec:analysis}

%

This section provides a comprehensive data analysis of the effect that various conditions had on the node temperatures for the Mustang cluster.  For brevity, we restrict attention to only the Mustang cluster.  A similar analysis was performed for the other compute machines in Room 341, but as mentioned previously Mustang was the most problematic and interesting.  The presentation here contains several incremental analyses that follow a chronological account of the data analysis as it occurred in practice from August 2013 through June 2014.  Thus, Section~\ref{sec:baseline} first discusses the analysis of only the data from a {\em baseline} experiment that was conducted on 8/7/2013 prior to any cooling changes.  Section~\ref{sec:temp_effect} then discusses the analysis of some subsequent cooling changes that were made during the next few months.  The new Wolf cluster was installed in the room in February 2014; Section~\ref{sec:wolf_effect} examines the effect of adding the new Wolf cluster to the room and unravels the cause of an overheating epidemic.  The results of Section~\ref{sec:wolf_effect} caused a revision to the benchmark HPL program, which is discussed in Section~\ref{sec:new_HPL}.  Some suspicious results in Section~\ref{sec:wolf_effect} prompted the examination of the effect of the removal of trays within the rack and rack doors in Section~\ref{sec:doors_trays}.  Finally, Section~\ref{sec:predictions} provides predictions of the current state-of-the-machine.


\vspace{-.2in}
\subsection{Baseline State-of-the-Machine}
\vspace{-.1in}
\label{sec:baseline}

The model in (\ref{eq:node_model}) was fit using only the spatial intercept $\beta_0$ to data from a baseline experiment that was conducted on 8/7/2013 prior to any cooling changes.  That is, any $x_j$ covariates that we ultimately investigate in the following sections remained at fixed values here.  The posterior mean of $\varphi$ was 0.9998, indicating a strong dependence in the overall temperature level between neighbors.  The posterior mean of $\blambda$ was $\hat{\blambda} = [0.298 \:,\: 0.241 \:,\: 0.177 \:,\: 0.194 \:,\: 0.086 \:,\: 0.002 \:,\: 0.002]$ which can be used directly to assess the relative strength of dependence between the various neighbor types, however, correlation is a more intuitive and reliable measure for this \cite{Wall04}.  The correlations between each neighbor type, resulting from $\bQ_j$ using $\hat{\blambda}$ (averaged over the correlations between all neighbors of the respective type), are
\vspace{-.2in}\bdm
[0.882 \:,\: 0.870 \:,\: 0.846 \:,\: 0.848 \:,\: 0.756 \:,\: 0.561 \:,\: 0.557].  
\vspace{-.2in}\edm
Thus, there is strong dependence between neighbor types (1 and 3), and (2 and 4), the close proximity horizontal and vertical directions, respectively.  However, there is much less dependence between neighbor types 6 and 7 (i.e., across aisles).  The posterior mean value of $e^{-\theta}$ was 0.955, indicating a correlation of 0.955 between two observations (from the latent GP $Z_s$ in (\ref{eq:meta_delta})) one minute apart on the same node.

\begin{figure}[t!]
  \vspace{-.2in}
  \centering
  \caption{Upper $95\%$ CBs for the maximum temperatures achieved while running HPL continuously for one day.}
  \vspace{-.1in}
  \label{fig:baseline_state_cop}
  \includegraphics[width=.9\textwidth]{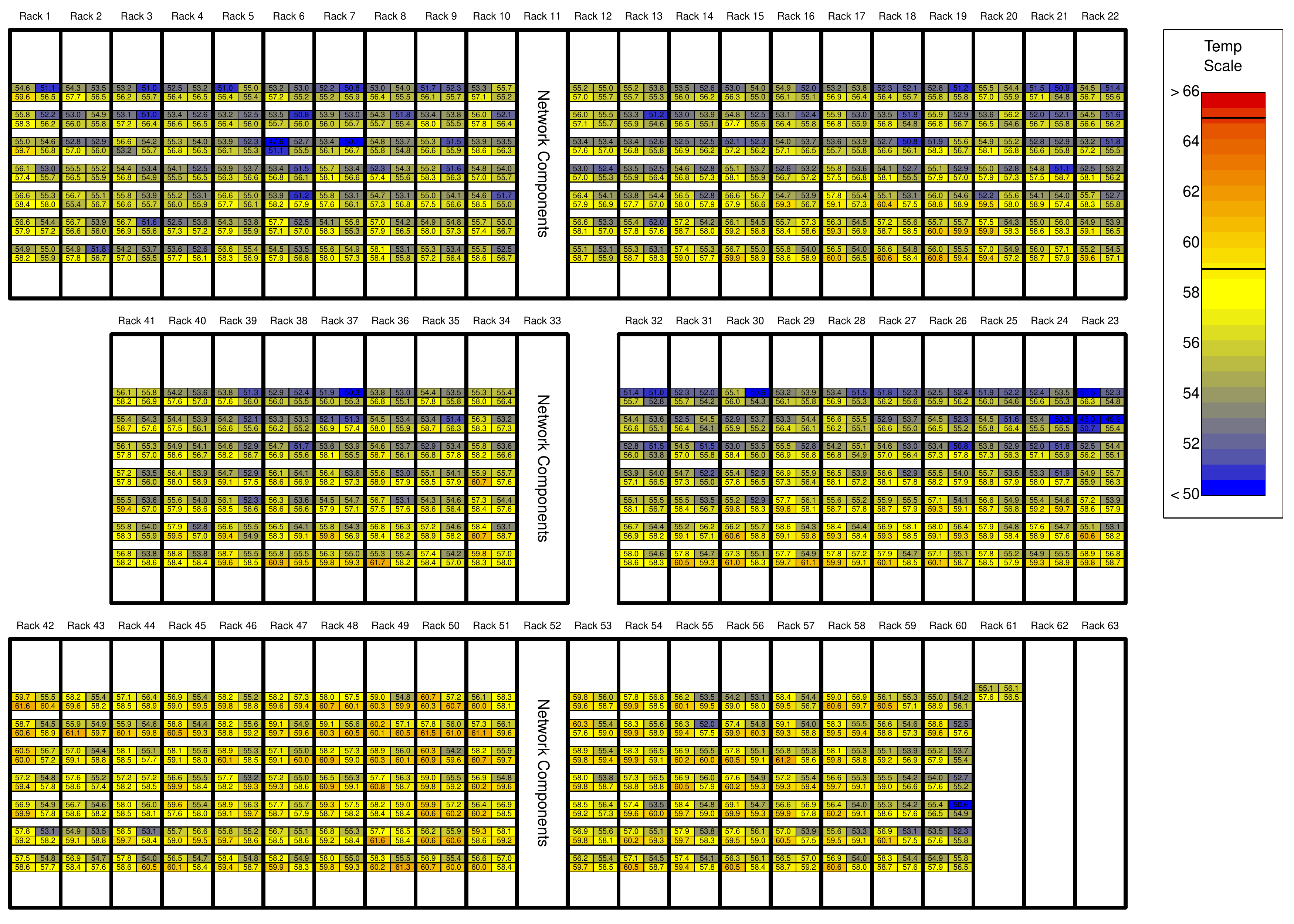}
\vspace{-.15in}
\end{figure}

The state-of-the-machine is defined to be an upper $95\%$ credible bound (CB) for the maximum temperature achieved (over all nodes) while running HPL continuously for one day.  This is obtained by producing a posterior predictive sample, $m=1,\dots,M$, of the values of temperatures $y_m(s,t)$ for another 24 hours using a dense time grid (i.e., every minute).  And then, for the $m\th$ posterior sample, extract the maximum temperature over $s$ and $t$; denote this maximum $Y^*_{m}$.  The state-of-the-machine is provided by $Y_{0.95}$, the $95\th$ percentile of the $Y^*_{m}$.  The state-of-the-machine under baseline room conditions is $63.5^\circ\:$C, indicating little chance of HPL producing a high temperature ($65^\circ\:$C) alert.  It is also helpful to complement this overall bound (for the maximum temperature of any node) with bounds for the individual node maxima, to locate hot spots.  That is, for the $m\th$ posterior sample, for each $s$, obtain the maximum of the $y_m(s,t)$ over $t$; denote these maxima $Y^*_{s,m}$.  A 95\% upper credible bound for the maximum achieved by node $s$ is the $95\th$ percentile of $Y^*_{s,m}$, denoted $Y_{s,0.95}$.  Figure~\ref{fig:baseline_state_cop} provides a graphical display of the $Y_{s,0.95}$.

\vspace{-.2in}
\subsection{Effect due to Cooling Changes in the Room}
\vspace{-.1in}
\label{sec:temp_effect}

After the baseline experiment, conducted on on 8/7/2013, and prior to 01/08/2014 three changes were made to the room.
Specifically, (i) four of the 18 CRAC units that provide cool air into the room were turned off.  Also, (ii) several cooling tile (i.e., the perforated tiles in the floor) changes were made to allow more airflow to certain locations of Mustang.  Finally, (iii) the upper band of the temperature controls (that govern the hottest air that the CRAC units can supply) was increased.  It was widely believed that increasing the upper temperature band would have little effect on node temperatures since the supply air from the CRAC units would be much more sensitive to the lower band temperatures (which remained unchanged).  For a detailed discussion of the supply temperature controls, see \citeasnoun{Michalak15}.

\begin{figure}[t!]
\vspace{-.1in}
  \centering
  \caption{Posterior mean of $\beta_1$:  The effect on temperature for each node after adjustment of cooling tiles, CRAC units, and upper temperature band.}
  \label{fig:beta1_mean}
  \vspace{-.1in}
  \includegraphics[width=.9\textwidth]{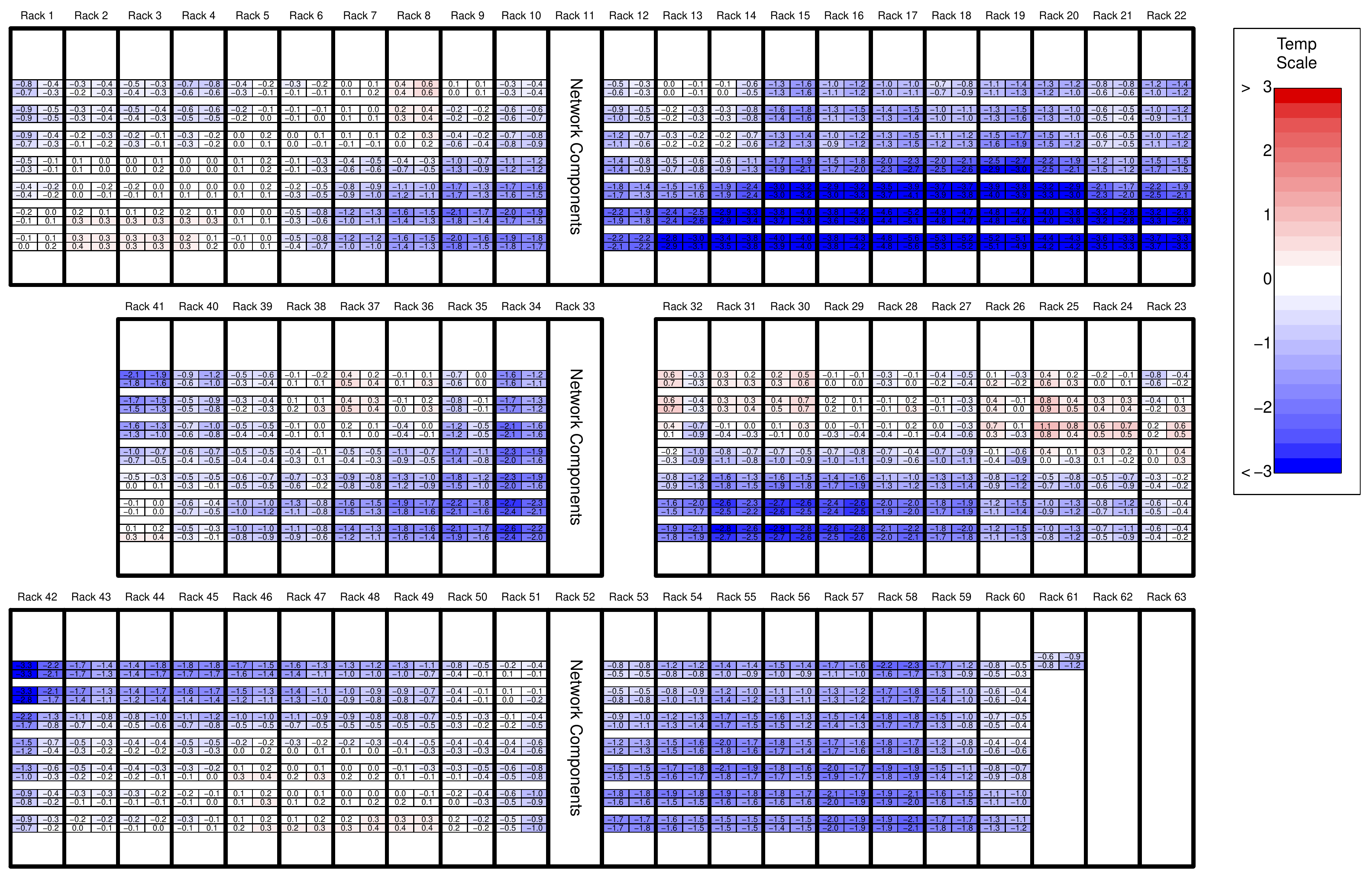}
  \vspace{.1in}
\end{figure}

\begin{figure}[t!]
\vspace{-.1in}
  \centering
\caption{Upper $95$\% CBs for $\beta_1$.}
  \label{fig:beta1_95}
  \vspace{-.1in}
  \includegraphics[width=.9\textwidth]{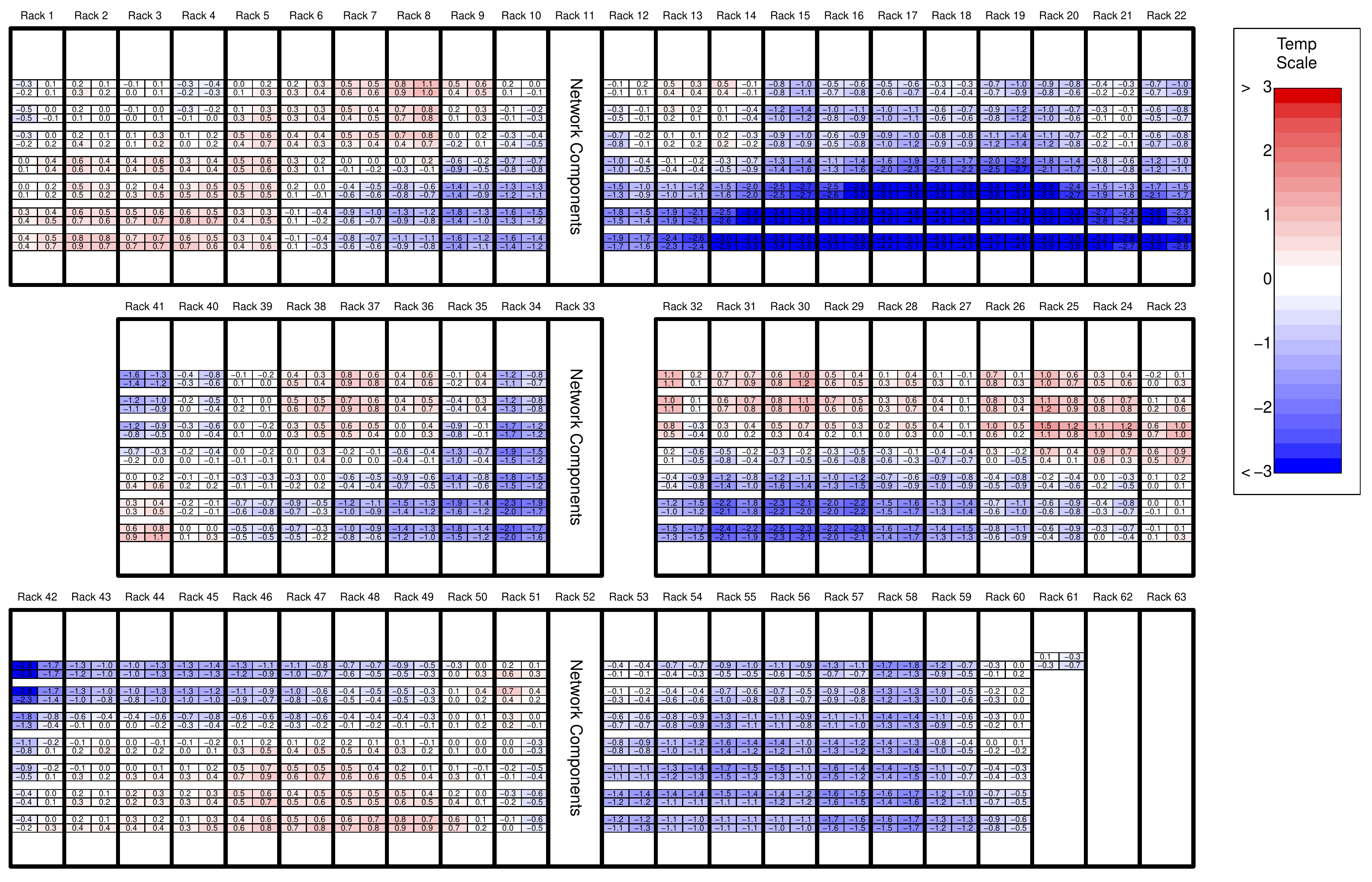}
\vspace{-.1in}
\end{figure}

Ideally all three of the factors listed above would be studied with separate experiments, however, getting in on a DST to obtain experimental data proved more difficult than anticipated.  Thus, all three effects listed above were confounded in this study.  The model in (\ref{eq:node_model}) was used to fit the data with only one covariate $x_1$ equal to 1 or 0 to indicate whether or not the CRAC unit/cooling tile/upper band changes had been made.  Figure~\ref{fig:beta1_mean} displays the resulting posterior mean of this effect ($\beta_1$) for each node, indicating that the cooling changes did not make a substantial increase in node temperatures anywhere.  The largest estimated increase for any node was 1.1$^\circ\:$C.  In fact, these changes had a negative effect overall on node temperatures (i.e., $-1^\circ\:$C averaged across nodes). Some effects were as low as $-5^\circ\:$C in the vicinity of Racks 15-20, where some of the perforated cooling tiles were added.
Figure~\ref{fig:beta1_95} displays 95\% upper CBs for the spatial effects, suggesting that the {\em largest} that the effect might plausibly be for any given node is about 1.6$^\circ\:$C and that many of the negative effects are significant.
Thus it is safe to conclude that the shutting down of four of the CRAC units did not have a substantial warming effect on the Mustang cluster.  Surprisingly, the combination of events (most likely the cooling tile changes) had a significant {\em cooling} affect on many of the nodes.

\vspace{-.2in}
\subsection{Nodes Started Overheating}
\vspace{-.1in}
\label{sec:wolf_effect}

In late February 2014, around the time the new Wolf cluster was installed in the room, reports of Mustang nodes overheating ($> 65^\circ\:$C) began coming in daily.  Some nodes were approaching $70^\circ\:$C.  The prominent theory was that the airflow around Mustang must have been substantially affected by the addition of Wolf.  The heavy majority of jobs that caused the critically high node temperatures belonged to three users.  It was determined that all three of the users had been running HPL jobs for performance testing purposes.  However, the parameters of the HPL job they were using had been ``optimized'' to provide maximum performance (Tflops/second).  For convenience we subsequently refer to this optimized version of HPL as HPL2 and the benchmark ({\em default} HPL) being previously used by the cooling team as HPL1.  An experiment was run on 04/09/2014 to assess the state-of-the-machine and attempt to uncover the cause of the overheating nodes.  It was considered unlikely at that time that HPL2 could be the cause of a $\sim10^\circ\:$C increase in node temperatures.  However, HPL1 and HPL2 were both run (for 30 minutes each) during the experiment.  Thus, at this point there are now three covariates in the model: $x_1$ - previous cooling changes effect (0 or 1), $x_2$ - presence of wolf cluster (0 or 1), and $x_3$ - Running HPL2 instead of HPL1 (0 or 1).

\begin{figure}[t!]
\vspace{-.1in}
  \centering
  \vspace{-.1in}
  \caption{Posterior mean effect due to the installation of Wolf (and any room changes that happened along with it).}
  \label{fig:wolf}
  \vspace{-.1in}
  \includegraphics[width=.9\textwidth]{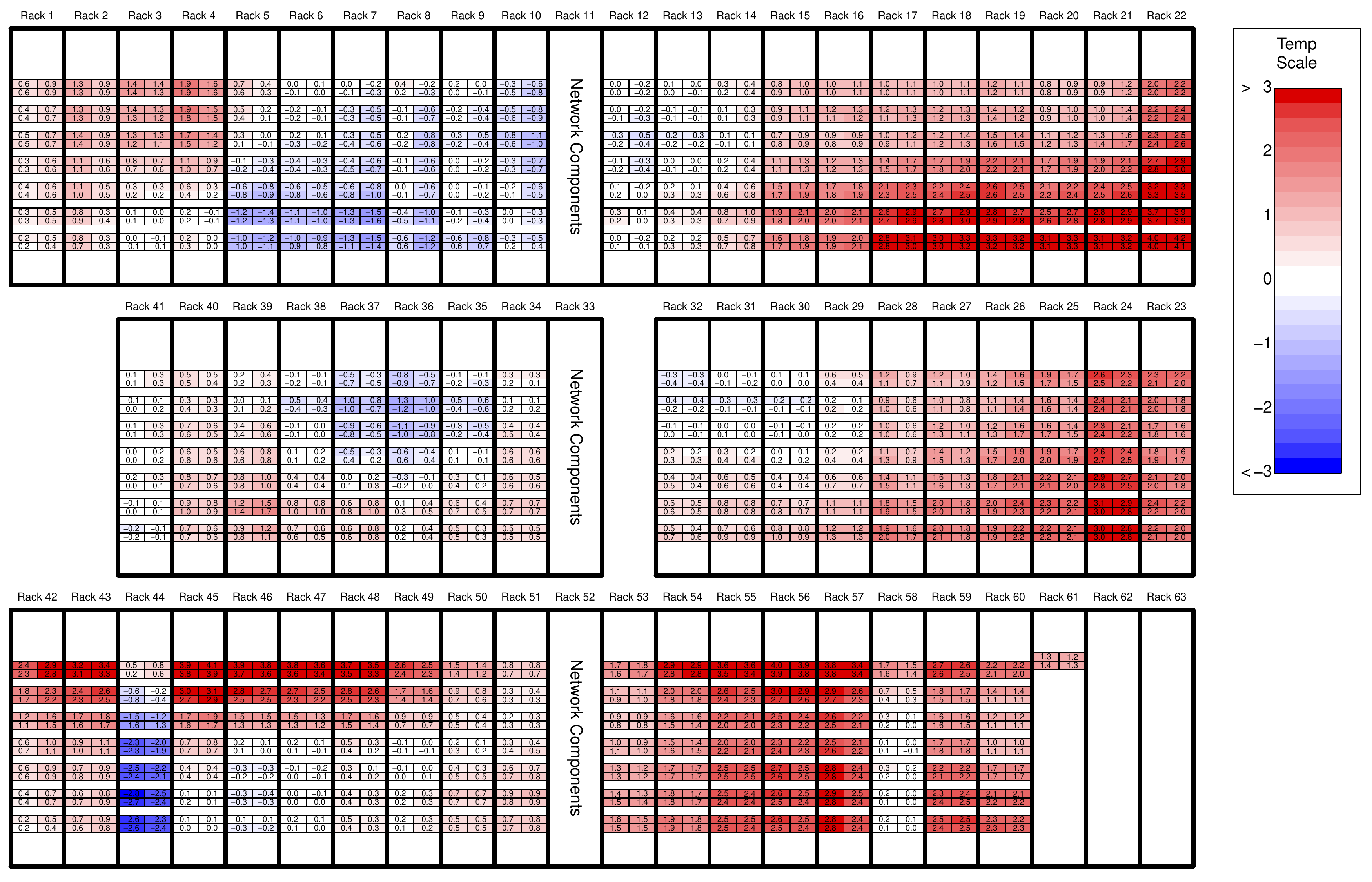}
\vspace{-.15in}
\end{figure}

\begin{figure}[t!]
\vspace{.1in}
  \centering
\caption{Posterior mean effect due to HPL2 being run instead of HPL1.}
  \label{fig:hpl2}
  \vspace{-.1in}
  \includegraphics[width=.9\textwidth]{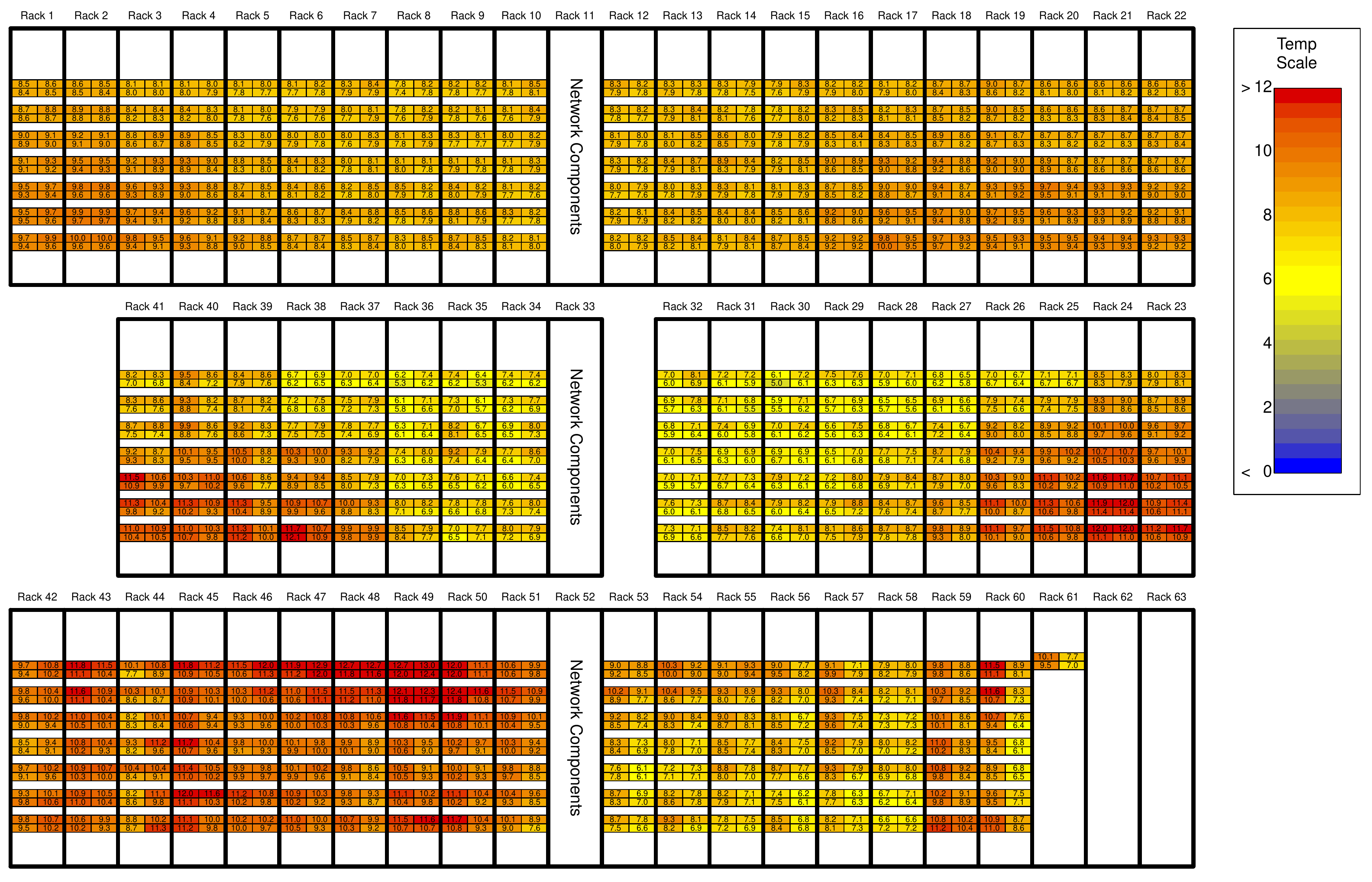}
\vspace{-.16in}
\end{figure}

Figure~\ref{fig:wolf} displays the effect that the addition of the Wolf cluster (and the airflow changes that came along with it) had on the Mustang node temperatures.  On average (across nodes) this affect is not large, but in some node locations it is close to 4$^\circ \:$C.  Still, considering the baseline individual bounds in Figure~\ref{fig:baseline_state_cop} (that had remained qualitatively unchanged by the cooling changes thus far), it is not possible that Wolf was causing the $\sim 10^\circ\:$C increases to node temperatures that were being observed in practice.  Interestingly, the effect of Wolf appeared to be abruptly different, spatially, for Racks~44~and~58.  These racks appeared to be running about 3$^\circ \:$C cooler than their respective neighboring racks.  After some investigation, it was discovered that the operations team had removed the support trays from these racks and left their doors open at some time prior to the experiment.  They had speculated that this may have an effect on node temperature.  It turns out that it does; more on this in Section~\ref{sec:doors_trays}.

Figure~\ref{fig:hpl2} displays the posterior mean effect on node temperature due to HPL2 being run instead of HPL1.  This plot basically tells the overheating story in a nutshell: HPL2 runs 8.5$^\circ \:$C hotter on average and as much as 13$^\circ \:$C hotter on some nodes than HPL1.  While the mystery of overheating nodes was solved, the prior understanding was that HPL was the most computational intensive job that a cluster might see.  But it was not understood that there are substantially different shades of HPL.  This prompted the cooling team to rethink the appropriateness of the current HPL1 benchmark.

\vspace{-.21in}
\subsection{A New Benchmark HPL}
\vspace{-.11in}
\label{sec:new_HPL}

\begin{figure}[t!]
\vspace{-.1in}
  \centering
  \caption{(a) Densities of the $95\th$ percentile of the hottest 40 users (magenta) versus HPL1 (blue).  Mean difference is about $3^\circ\:$C. (b) Density of the pairwise differences (by node) between the $95\th$ percentile of the hottest 40 users to HPL1.}
    \label{fig:hpl_vs_extreme}
    \begin{subfigure}[b]{.45\textwidth}
      \centering
\vspace{-.1in}
      \caption{}
\vspace{-.1in}
      \includegraphics[width=.88\textwidth]{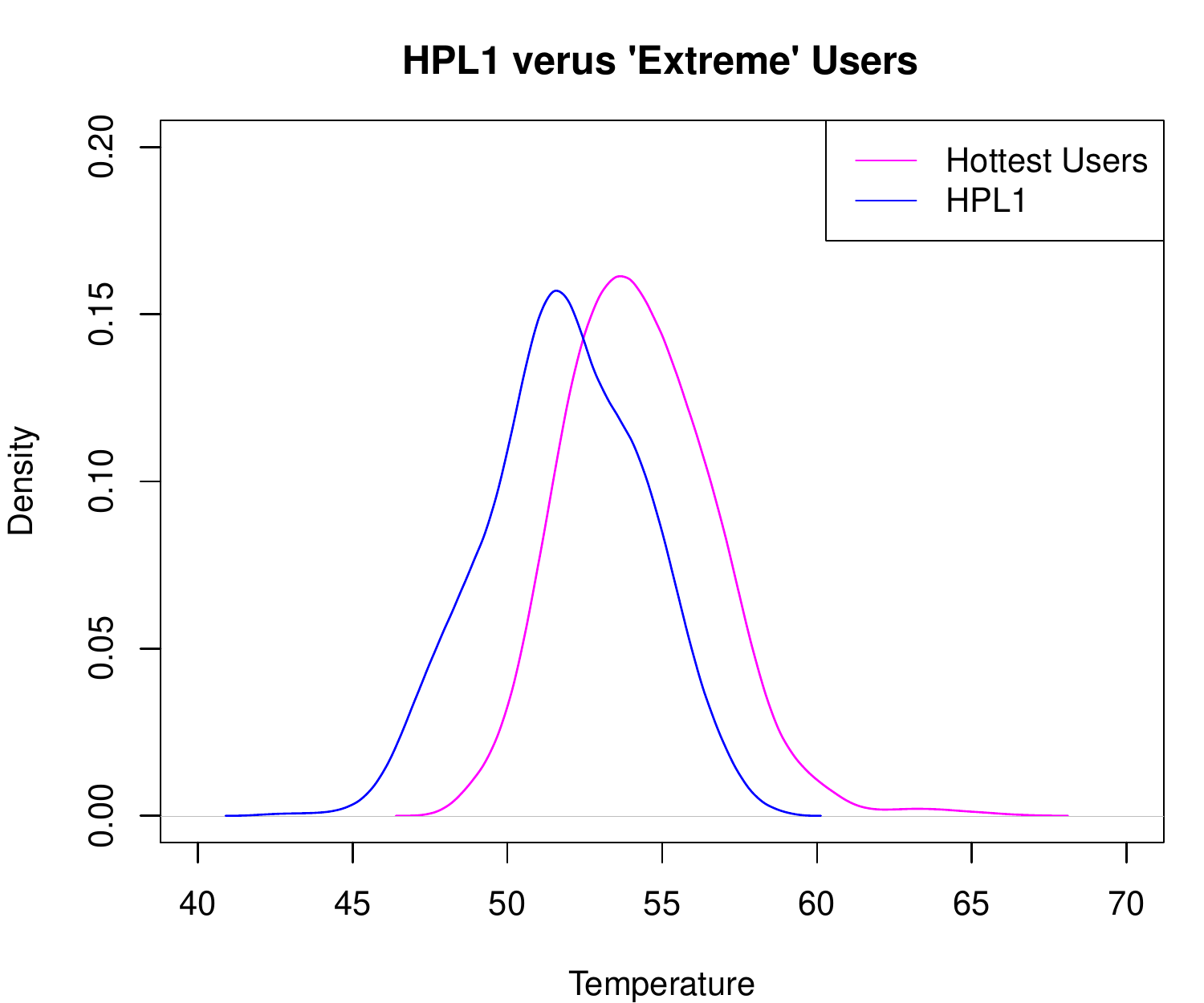}
    \end{subfigure}
    \begin{subfigure}[b]{.45\textwidth}
      \centering
\vspace{-.1in}
      \caption{}
\vspace{-.1in}
      \includegraphics[width=.88\textwidth]{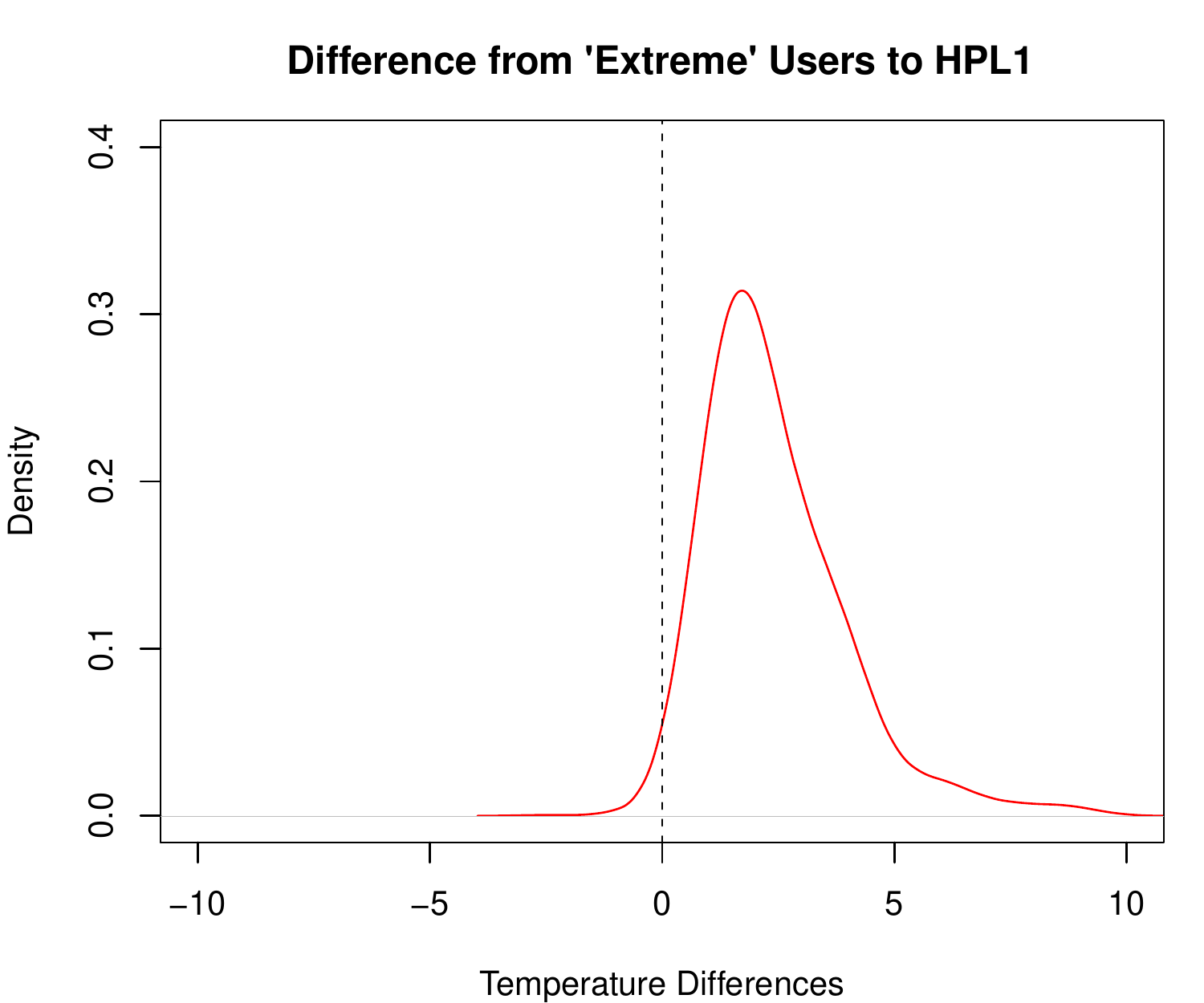}
    \end{subfigure}
\vspace{-.0in}
\end{figure}

Should HPL2 be used as the benchmark instead of HPL1, or is HPL2 too conservative?  What do ``extreme'' node temperatures on Mustang look like under regular, production use?  Temperature records during regular use are recorded $\sim$every hour in a database for most of the LANL machines (including Mustang).  Thus, these records were used, along with job history records for the previous four months to identify the 40 ``hottest'' users of Mustang.  An extreme-user-dataset was created that had all temperature readings for any node that ran one of the 40 hottest users' jobs.  The empirical $95\th$ percentile of the extreme-user-dataset was calculated by node and the resulting density (across nodes) is displayed in Figure~\ref{fig:hpl_vs_extreme}(a).  For comparison, the density (across nodes) for the empirical $95\th$ percentile temperature achieved while running the HPL1 benchmark during the same time frame is also displayed as the dashed curve in Figure~\ref{fig:hpl_vs_extreme}(a).  The density of the pairwise differences (between nodes) of these $95\th$ percentiles (extreme users minus HPL1) is displayed in Figure~\ref{fig:hpl_vs_extreme}(b).  It is apparent that the HPL1 benchmark was not intensive enough.  Typical use by the hottest users of Mustang results in node temperatures that are about $3^\circ\:$C hotter on average than that of HPL1.   However, it is also clear that HPL2 (which is $8.5^\circ\:$C hotter than HPL1 on average) would be far too conservative.

\begin{figure}[t!]
\vspace{-.12in}
  \centering
  \caption{(a) Average (across sampled nodes) HPL temperature difference from HPL1 achieved while varying N and Block Size. Surface estimated with smoothing splines. The setting for HPL1 is given by the magenta point. Block Size of 5 should provide $\sim 5^\circ\:$C increase. (b) Distribution of the pairwise differences (by node) between the $95\th$ percentile of the hottest 40 users to HPL1 and the distribution (over nodes) of HPL3 effect (relative to HPL1).}
  \label{fig:hpl_contour}
  \vspace{-.15in}
    \begin{subfigure}[b]{.48\textwidth}
      \centering
\vspace{-.0in}
      \caption{}
\vspace{-.05in}
      \includegraphics[width=.89\textwidth]{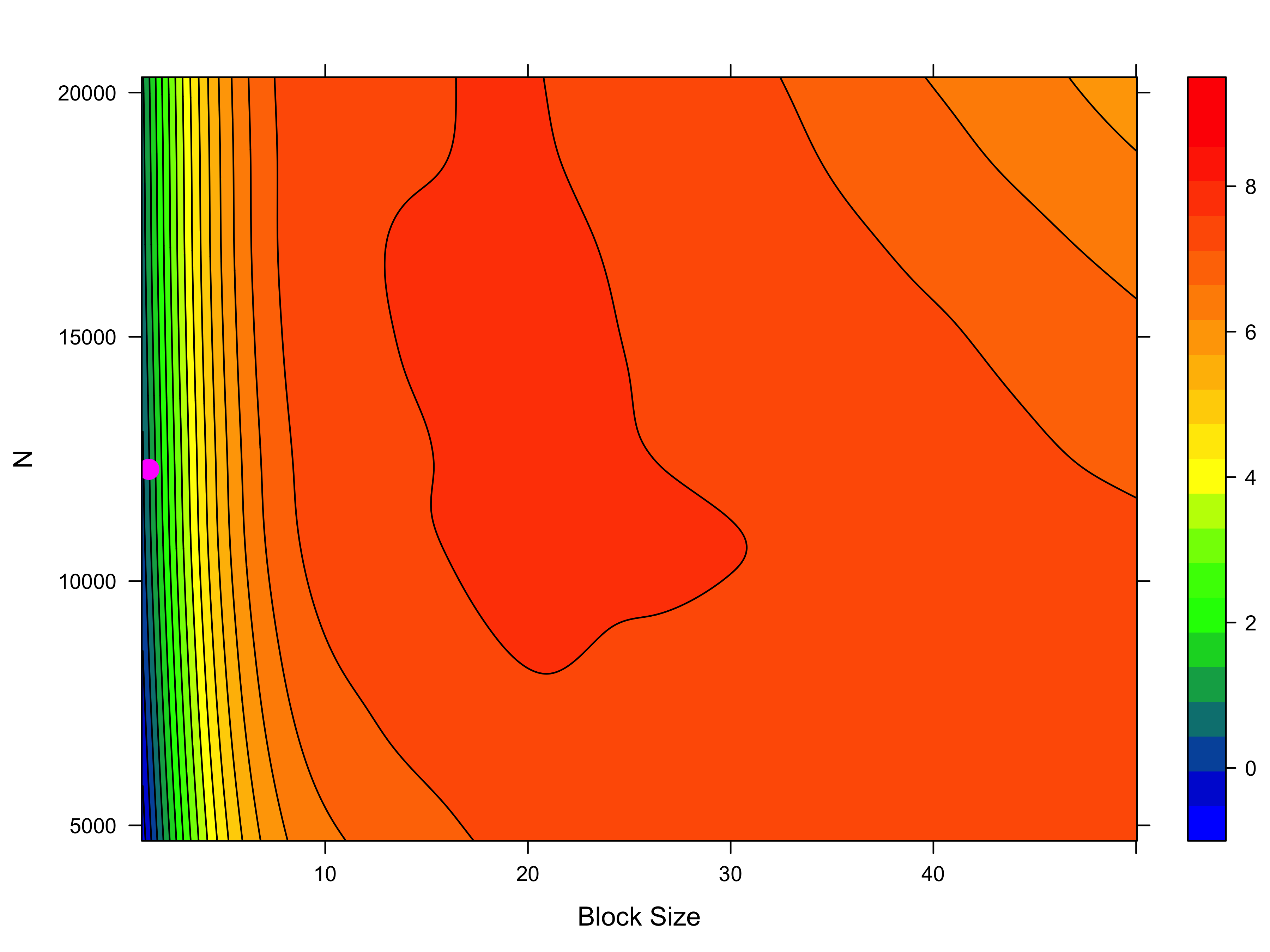}
    \end{subfigure}
    \begin{subfigure}[b]{.47\textwidth}
      \centering
\vspace{-.2in}
      \caption{}
\vspace{-.1in}
      \includegraphics[width=.9\textwidth]{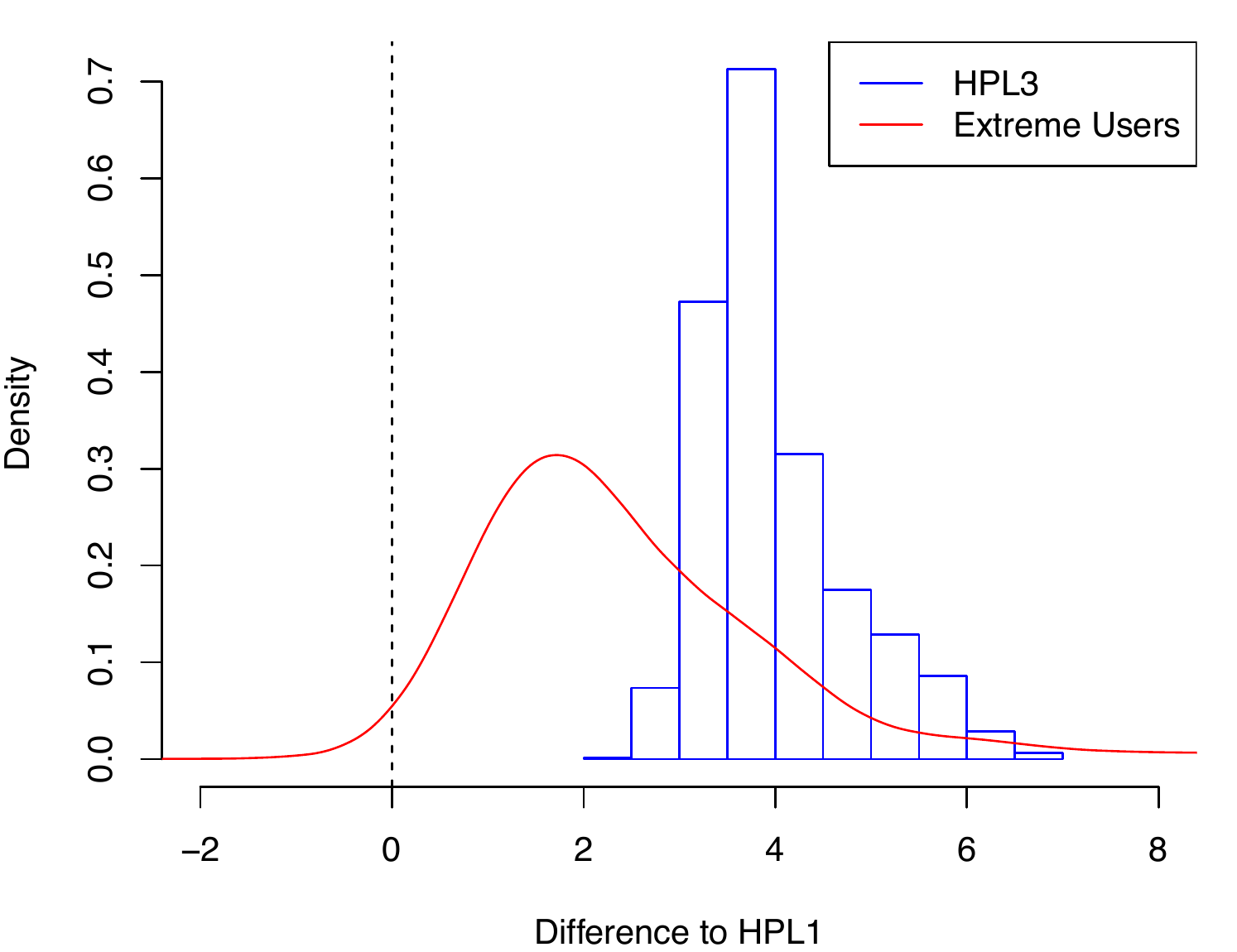}
    \end{subfigure}
\vspace{.02in}
\end{figure}

It was decided that a new benchmark should be constructed that more closely resembled the regular hottest users of Mustang.  Thus, an experiment was conducted varying the parameters of HPL, to identify the appropriate setting.  The two HPL parameters varied were $N$ - the size of the matrix, and {\em block size} - the size of the tasks sent out to each worker.  The design was chosen by a Latin-hypercube sample of 25 (with one run at the HPL1 settings, $N=12,800$ and {\em block size}$\;=1$).  A DST was not available to conduct this experiment, so many regular jobs (each consisting of the 26 HPL runs) were submitted to the standard queue to get coverage of the nodes.  In all, the 26 HPL settings were run on $\sim 900$ nodes with nearly uniform coverage over space.  The resulting response surface of the average temperature (across nodes), provided in Figure~\ref{fig:hpl_contour}(a), was estimated via an additive smoothing spline with a multiplicative interaction term using the generalized additive model framework \cite{Hastie90}.  The HPL1 design point with {\em block size}$\;=1$ and $N=12,800$ is plotted for reference.  Node temperature is not sensitive to $N$, but very sensitive to {\em block size}, particularly for small values of {\em block size}.

\begin{figure}[t!]
\vspace{-.12in}
  \centering
  \caption{Posterior mean effect due to HPL3 being run instead of HPL1.}
  \vspace{-.1in}
  \label{fig:hpl3}
  \includegraphics[width=.9\textwidth]{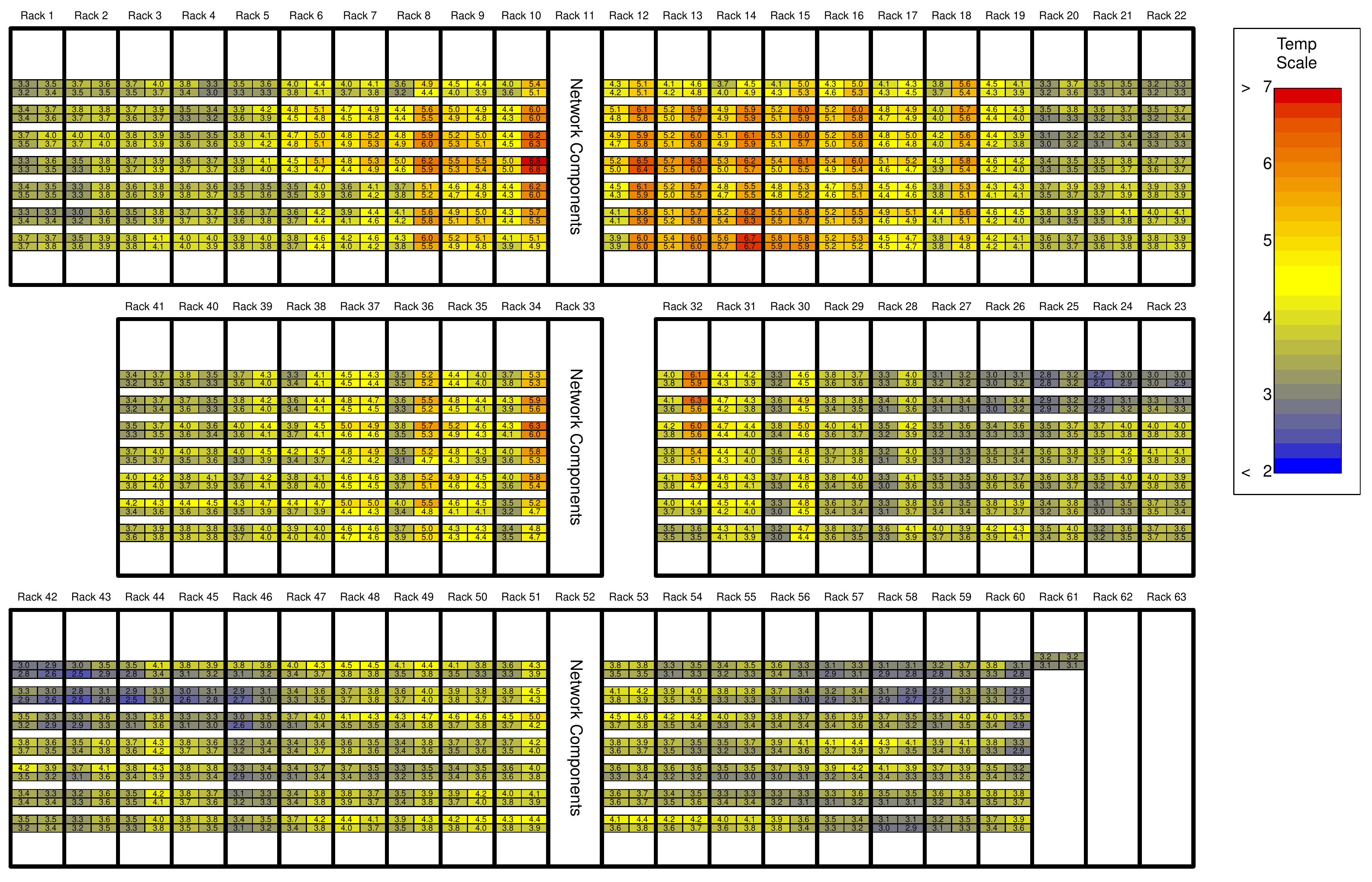}
  \vspace{-.16in}
\end{figure}

It was determined that HPL with {\em block size}$\;=4$ and $N=12,800$ (hereafter referred to as HPL3) would provide $\sim 4^\circ\:$C increase on average across nodes.  This setting was then run during the next DST (discussed in detail in Section~\ref{sec:doors_trays}) to confirm this finding on the entire Mustang cluster.  Figure~\ref{fig:hpl3} displays the posterior mean effect across nodes of running HPL3 instead of HPL1 after fitting the model in (\ref{eq:node_model}) with these new DST data.  Figure~\ref{fig:hpl_contour}(b) displays the density of the pairwise differences between the extreme users on Mustang to HPL1 again, along with a histogram of the posterior mean of the node effects due to HPL3.  HPL3 is if anything slightly conservative, as desired.

\vspace{-.21in}
\subsection{Effect of Removing Trays and Doors}
\vspace{-.11in}
\label{sec:doors_trays}

As mentioned in Section~\ref{sec:wolf_effect}, the effect of Wolf appeared to be abruptly different spatially for Racks~44~and~58 in Figure~\ref{fig:wolf} due to the fact that the operations team had removed the support trays from these racks and left their doors open.  This prompted the cooling team to investigate the effect of tray removal from a rack and that of opening doors.  Thus on 06/11/2014 an experiment was conducted to assess the affect of tray and door removal, and also to confirm the effect of the new benchmark HPL3.  As it requires a nontrivial amount of time to remove trays, the trays were removed prior to the experiment on all even numbered racks and remained out for the entirety of the experiment.  During the experiment the following conditions were tested: (i)  HPL1 with all doors closed and trays out of even racks, (ii) HPL1 with even numbered doors opened and trays out of even racks, and (iii) HPL3 with even numbered doors opened and trays out of even racks.  Only the even numbered rack doors where opened since the operations team decided it would be difficult logistically to open all doors at once.  Obviously, a more comprehensive factorial design would be preferred, but as before, only a very limited amount of time was available for experimentation on the Mustang cluster.  Even still, the effect of the removal of doors and trays can be estimated for each node due to the spatial correlation of effects in model (\ref{eq:node_model}).  After inspection of the effect plots, there was some clear interaction between tray removal and the HPL3 effects.
Thus, the present model includes the following covariates: $x_1$ - previous cooling changes effect (0 or 1), $x_2$ - presence of wolf cluster (0 or 1), $x_3$ - Running HPL2 instead of HPL1 (0 or 1), $x_4$ - Running HPL3 instead of HPL1 (0 or 1), $x_5$ - Trays in or out (0 or 1), $x_6$ - doors closed or open (0 or 1), and $x_7$ - Interaction of running HPL3 and trays out ($x_4 x_5$).  For clarification, in the present analysis, racks~44~and~58 also now use $x_5=x_6=1$ when using the  experimental data from the previous experiment 04/09/2014.  This was intentionally not the case in the analysis of Section~\ref{sec:wolf_effect} in order to preserve the natural order of analysis and discovery.

\begin{figure}[t!]
\vspace{-.1in}
  \centering
\caption{Posterior mean effect due to the removal of trays while running HPL3.  On average (across nodes) this affect is a $\sim$3.5$^\circ \:$C reduction.}
  \label{fig:trays}
\vspace{-.1in}
  \includegraphics[width=.9\textwidth]{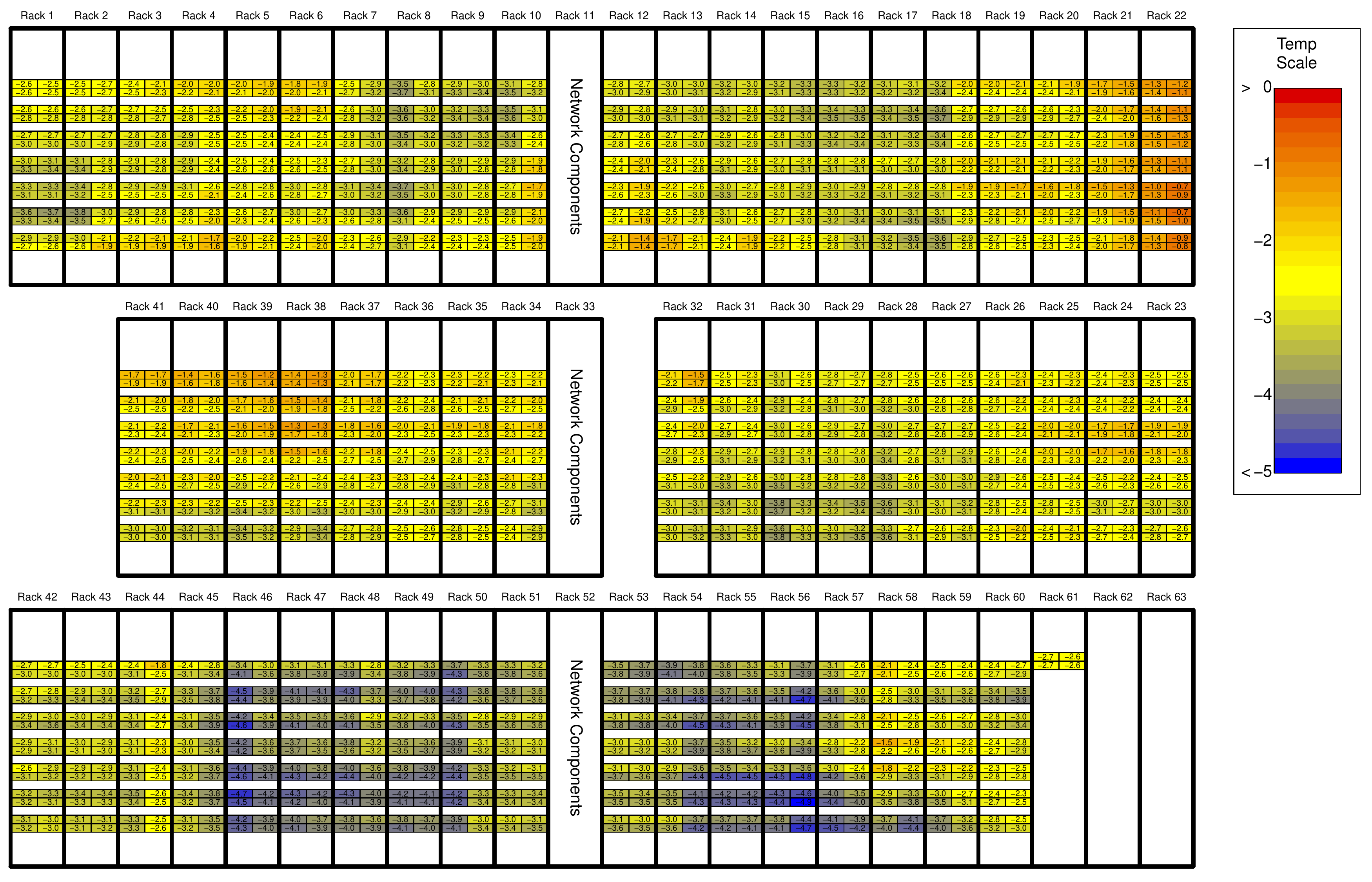}
\vspace{-.13in}
\end{figure}

Figure~\ref{fig:trays} displays the posterior mean effect due to tray removal while running HPL3.  This effect is $-2.8^\circ\:$C on average across nodes and as much as $-5^\circ\:$C for some nodes.  The posterior mean effect of opening doors is not nearly as pronounced (only $-0.1^\circ\:$C on average across nodes).  Based on 95\% lower CBs, the most beneficial the effect of opening doors could be for any given node is $-0.9^\circ\:$C.  Thus, the cooling benefit that was noticed on racks~44~and~58 in Figure~\ref{fig:wolf} must have come from tray removal.


\vspace{-.2in}
\subsection{Assessing the State-of-the-Machine}
\begin{figure}[t!]
  \vspace{-.13in}
  \centering
\caption{Upper 95\% CBs for the max node temperatures that would be achieved by running HPL3 continuously for one day with Wolf in the room, trays {\em out}, and doors closed.}
  \label{fig:SOR_63}
\vspace{-.1in}
\includegraphics[width=.9\textwidth]{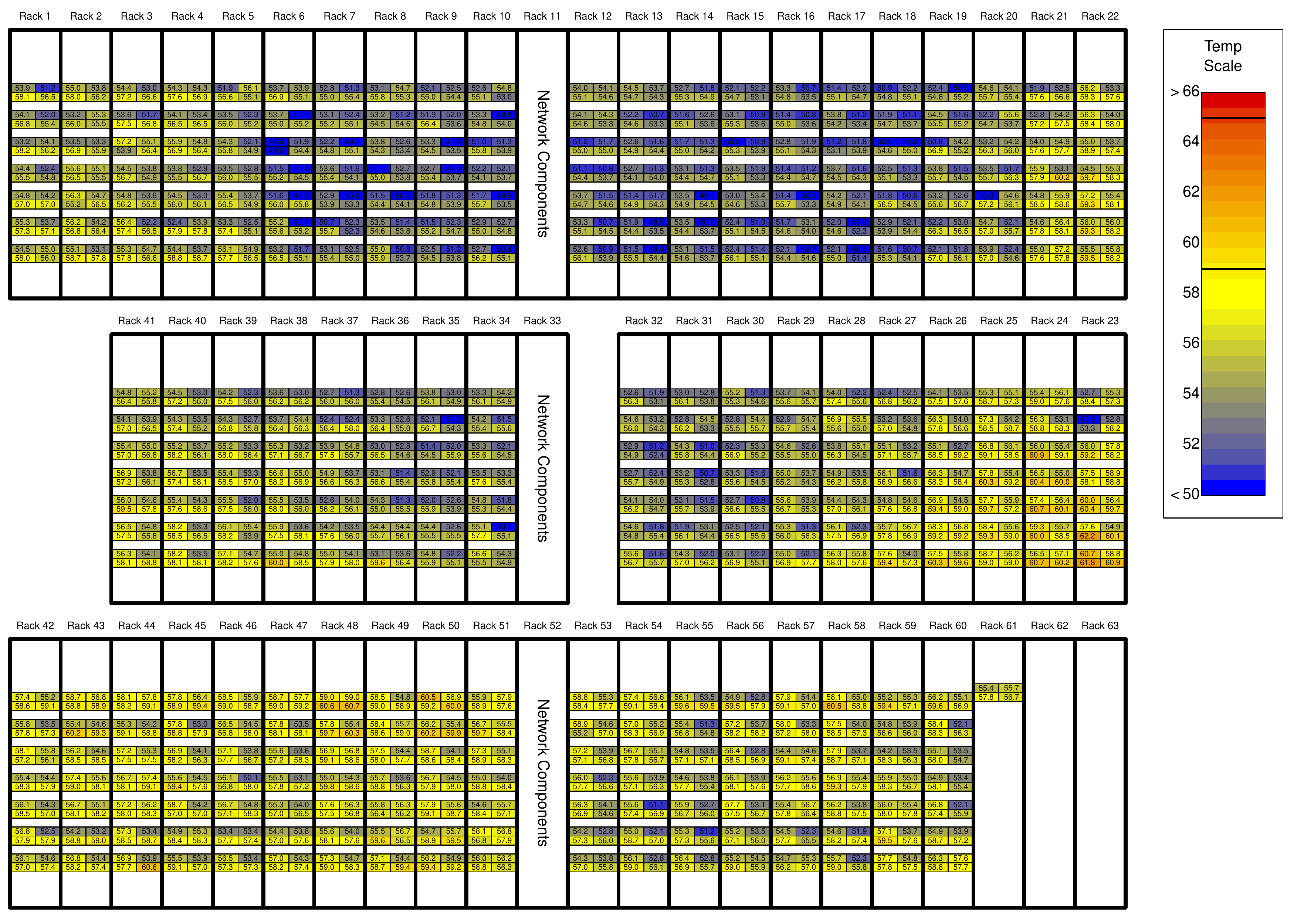}
  \vspace{-.16in}
\end{figure}

\vspace{-.1in}
\label{sec:predictions}
Finally, it is helpful to assess the current state-of-the-machine after changes and to highlight any potential problem areas of nodes.  The current state-of-the-machine with HPL3, Wolf in the room, trays {\em out}, and doors closed, using the posterior distribution resulting from the model fit in Section~\ref{sec:doors_trays} is $64.1^\circ\:$C.  This is only slightly hotter than the baseline state-of-the-machine ($63.5^\circ\:$C), still below the high ($65^\circ\:$C) threshold and well below the critical ($70^\circ\:$C) threshold.  Figure~\ref{fig:SOR_63} displays the corresponding individual upper 95\% CBs on the node maxima.  Compared to the those from baseline in Figure~\ref{fig:baseline_state_cop}, the predicted maximum temps are a bit hotter in some areas and lower in others, with the most problematic nodes now coming from Rack 23.

\vspace{-.2in}
\subsection{Assessing Model Assumptions}
\vspace{-.1in}
\label{sec:assumptions}

\begin{figure}[t!]
\vspace{.2in}
  \centering
  \caption{Normal, Normal$+$GPD, and nonparametric estimated marginal densities for $\delta$.}
    \label{fig:Norm_GPD_plots}
    \begin{subfigure}[b]{.49\textwidth}
      \centering
      \vspace{-.1in}
      \caption{Histogram and density estimates for $\delta$.}
      \vspace{-.15in}
      \includegraphics[width=.84\textwidth]{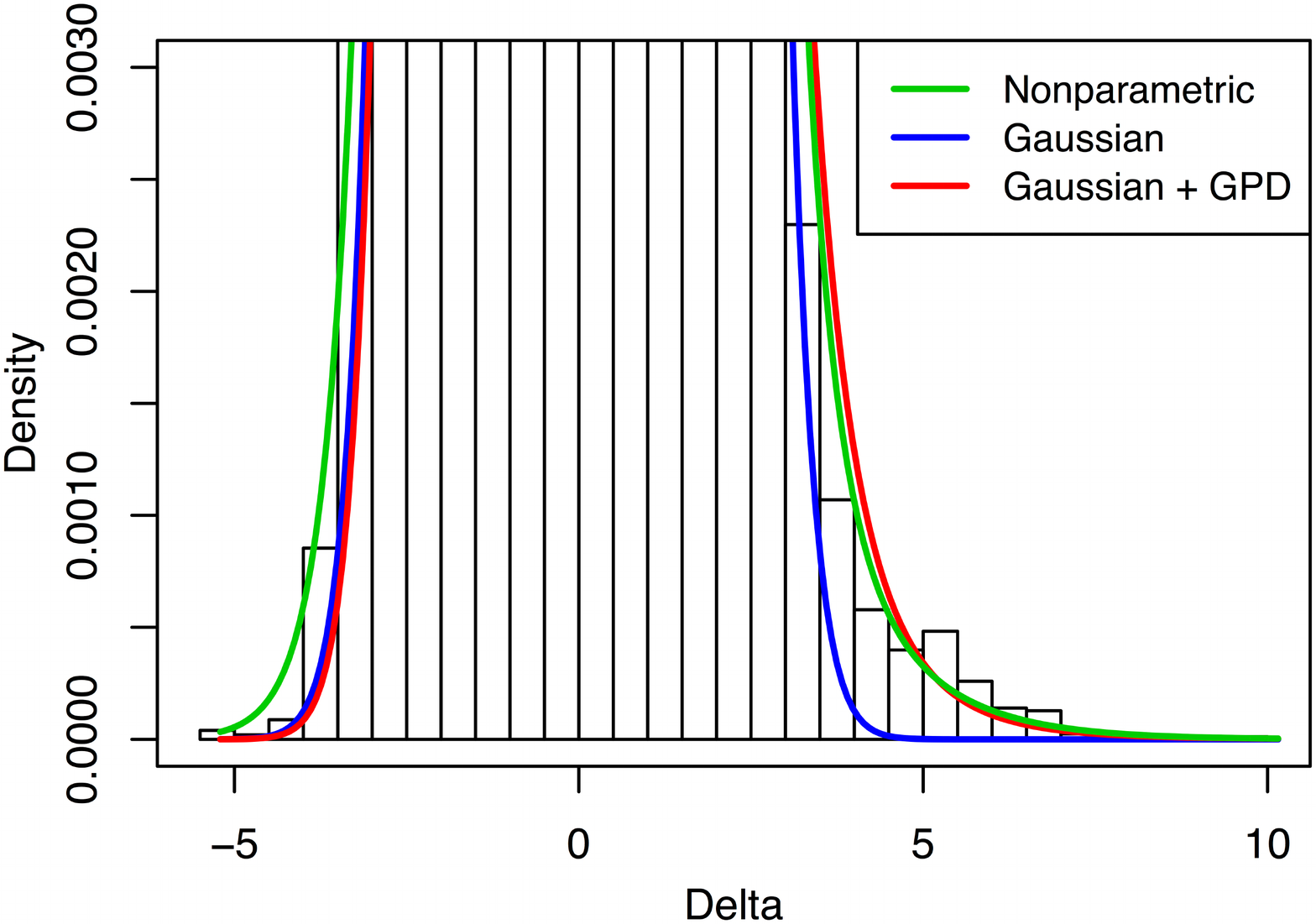}
    \end{subfigure}
    \begin{subfigure}[b]{.49\textwidth}
      \centering
      \vspace{-.1in}
      \caption{log-density estimates for $\delta$.}
      \vspace{-.15in}
      \includegraphics[width=.84\textwidth]{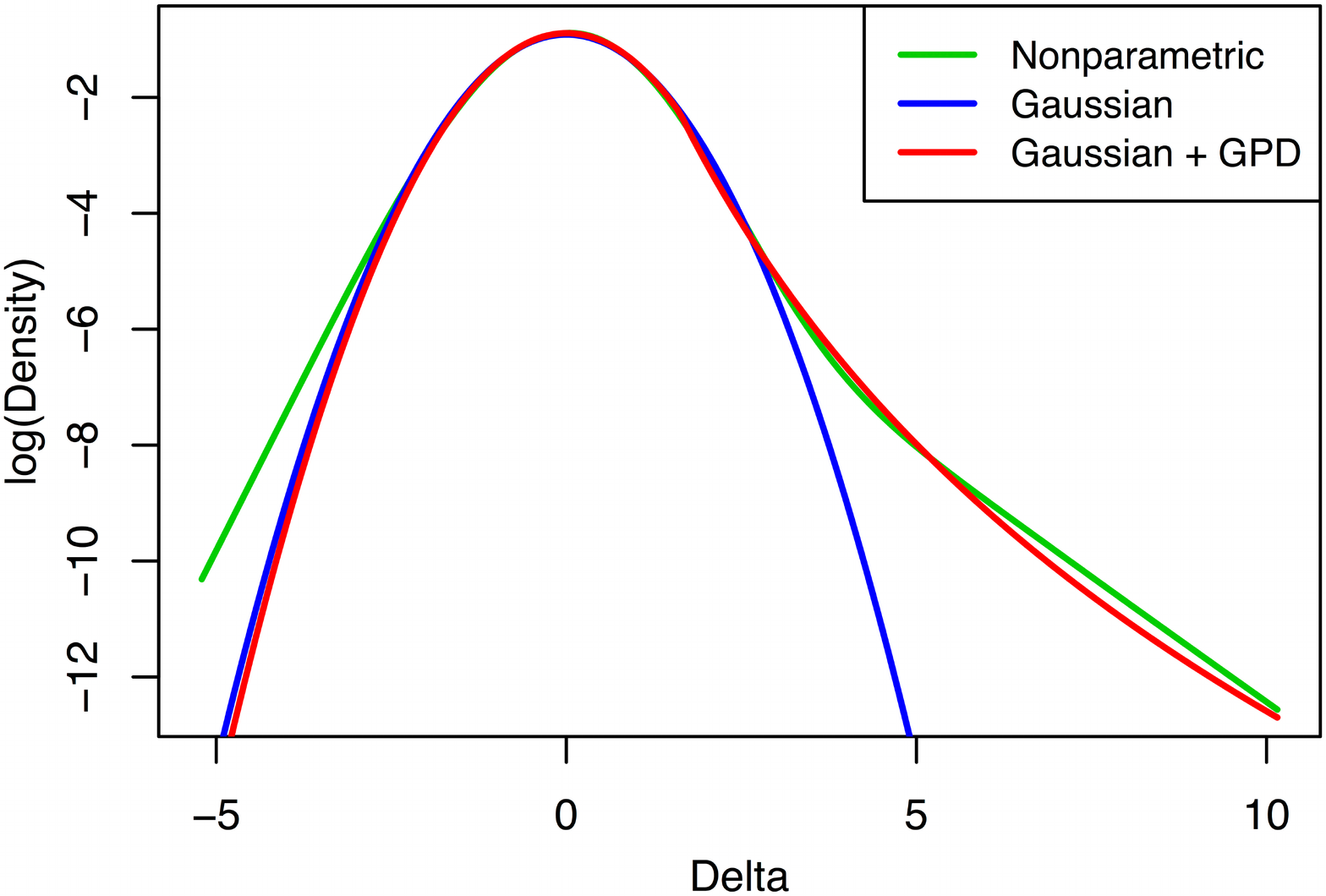}
    \end{subfigure}
    \begin{subfigure}[b]{.49\textwidth}
      \centering
      \vspace{-.0in}
      \caption{Gaussian Q-Q plot for $\hdelta$.}
      \vspace{-.15in}
      \includegraphics[width=.835\textwidth]{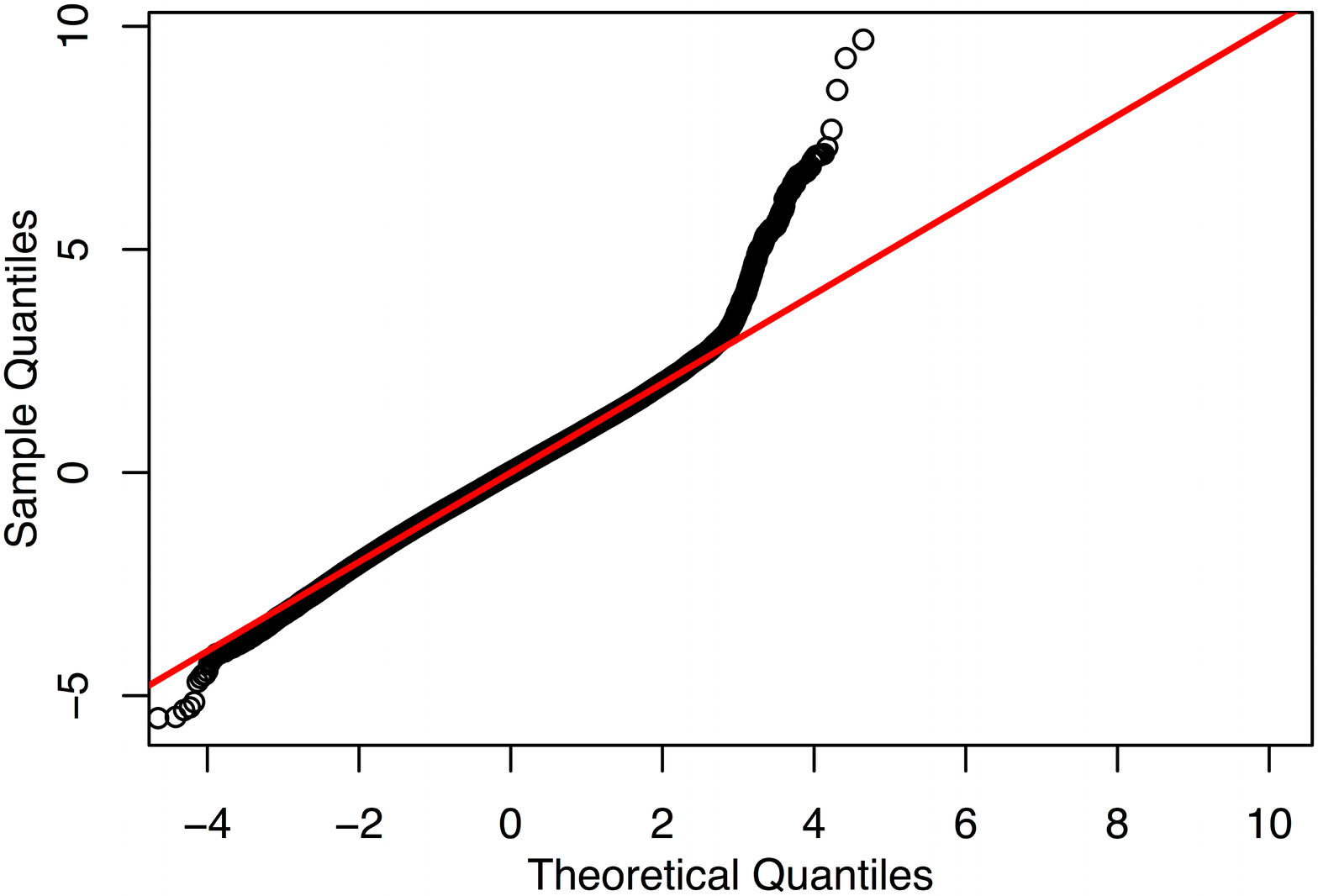}
    \end{subfigure}
    \begin{subfigure}[b]{.49\textwidth}
      \centering
      \vspace{-.0in}
      \caption{Gaussian $+$ GPD  Q-Q plot for $\hdelta$.}
      \vspace{-.15in}
      \includegraphics[width=.84\textwidth]{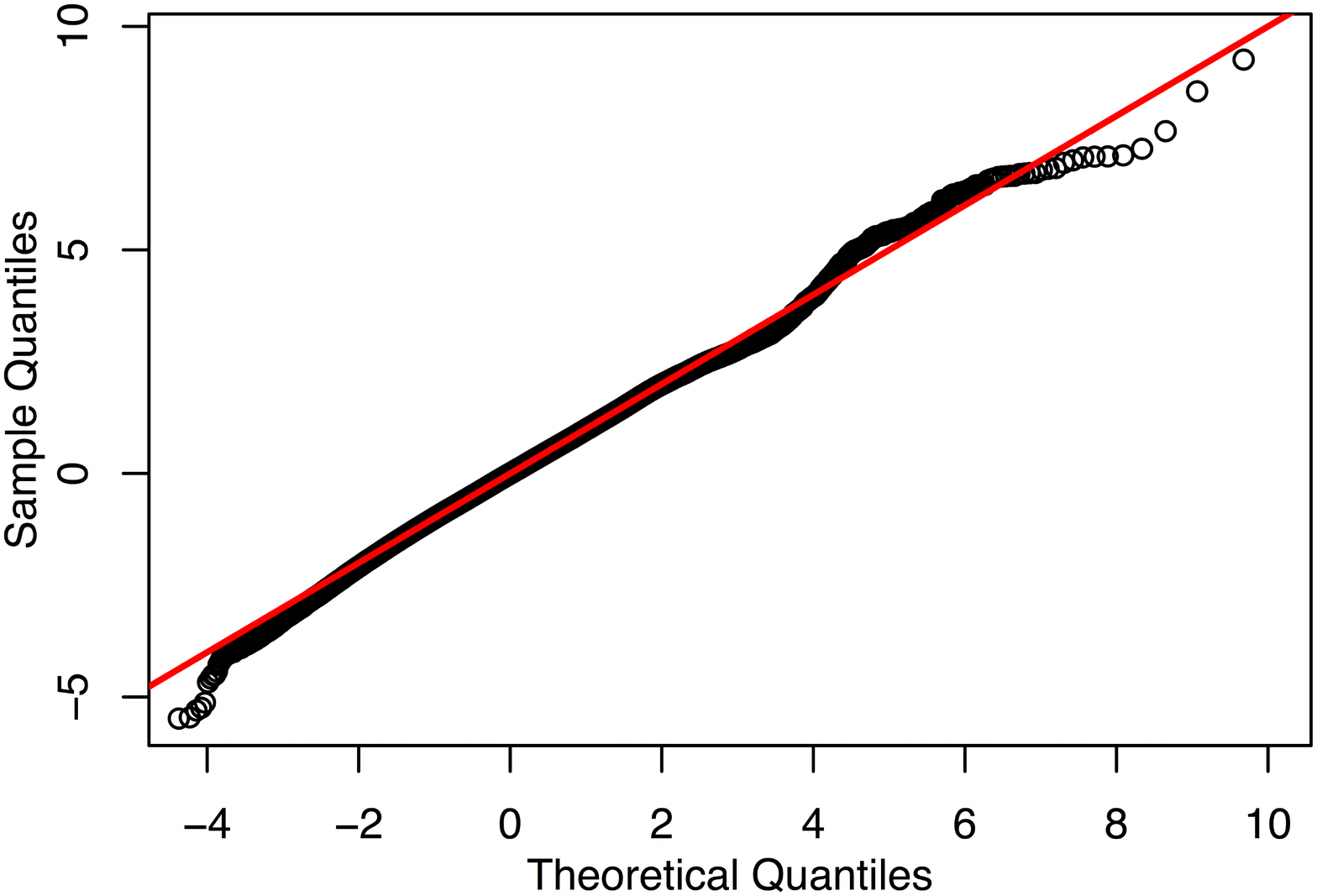}
    \end{subfigure}
\vspace{-.4in}
\end{figure}

It is prudent to assess some of the key modeling assumptions underlying the analysis and conclusions in the preceding sections.  In order to investigate some of these assumptions, the {\em fitted} (i.e., posterior mean) values $\hdelta$ were obtained.  Since the state-of-the-machine is heavily dependent on extremes, it is important to ensure that $\delta$ is accurately represented, particularly in the tail.  Figure~\ref{fig:Norm_GPD_plots} provides a histogram and normal Q-Q plot for the $\hdelta_s(t)$, for all values of $s$ and $t$.  If we had assumed a GP for each $\delta_s$, we would expect the marginal distribution of $\delta_s(t)$ to be $N(0,\upsilon^2)$, however, this is clearly not the case.  A nonparametric logspline density estimate \cite{Kooperberg91} is provided along with the Normal$+$GP density fit (and corresponding Q-Q plot) using the posterior mean values of $\upsilon^2=0.95$, $\kappa=1.66$, and $\xi=0.120$.

\begin{figure}[t!]
\vspace{-.1in}
  \centering
  \caption{Space-time dependence in $\delta$.}
    \label{fig:delta_dependence}
    \begin{subfigure}[b]{.49\textwidth}
      \centering
      \vspace{-.1in}
      \caption{Cor$(\hZ(s,t), \hZ(s,t-h))$ as a function of\\ \white{....} $\:$the time lag $h$.}
      \vspace{-.15in}
      \includegraphics[width=.88\textwidth,height=.19\textheight]{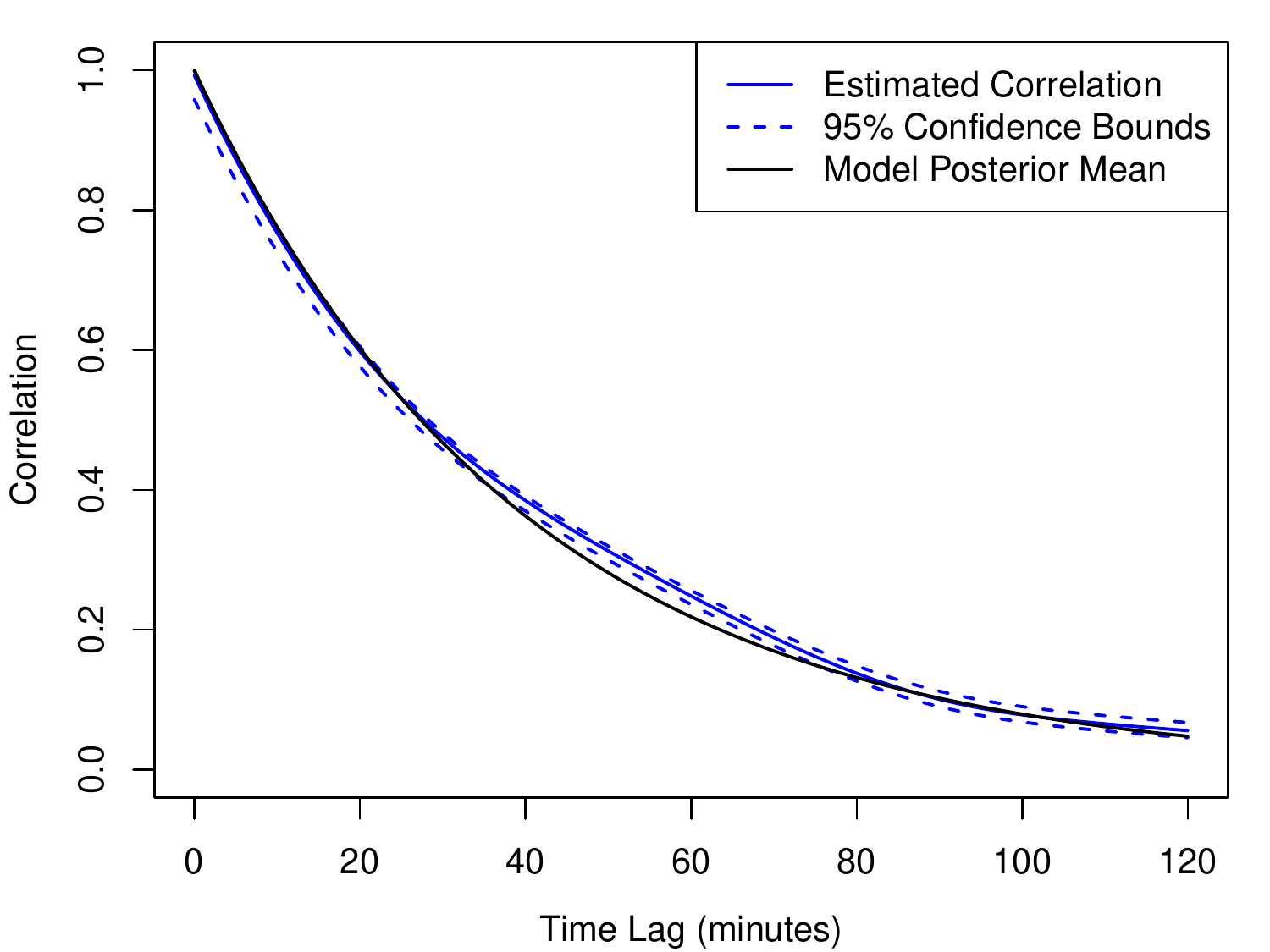}
    \end{subfigure}
    \begin{subfigure}[b]{.49\textwidth}
      \centering
      \vspace{-.1in}
      \caption{Tail dependence between time neighbors \\ \white{.....} $\hdelta(s,t)$ and $\hdelta(s,t-1)$.}
      \vspace{-.15in}
      \includegraphics[width=.88\textwidth,height=.19\textheight]{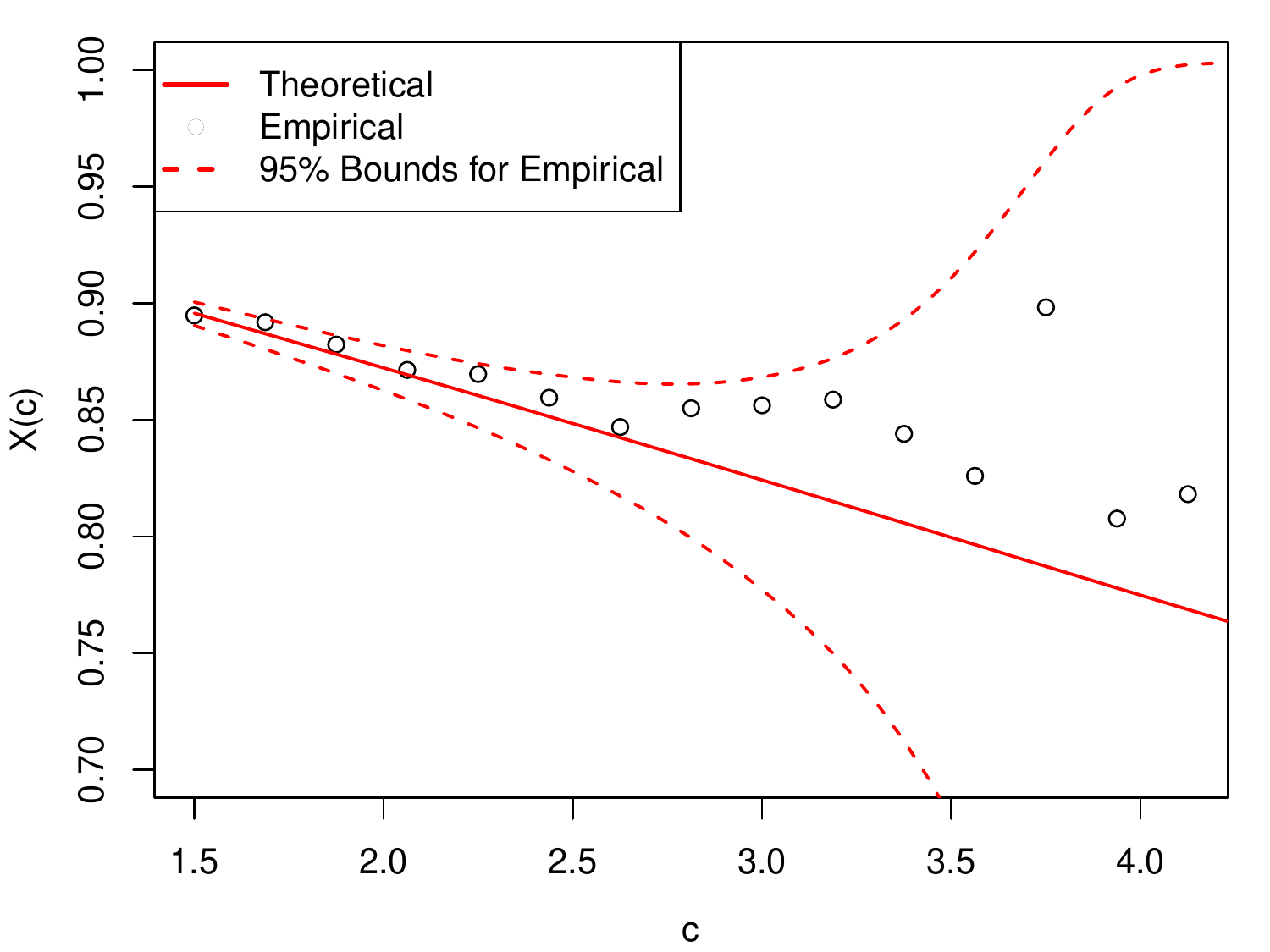}
    \end{subfigure}
    \begin{subfigure}[b]{.49\textwidth}
      \centering
      \vspace{.1in}
      \caption{Cor$(\hZ(s,t), \hZ(s',t))$ as a function of \\ \white{....} Euclidean distance $\|s-s'\|$.}
      \vspace{-.15in}
      \includegraphics[width=.88\textwidth,height=.19\textheight]{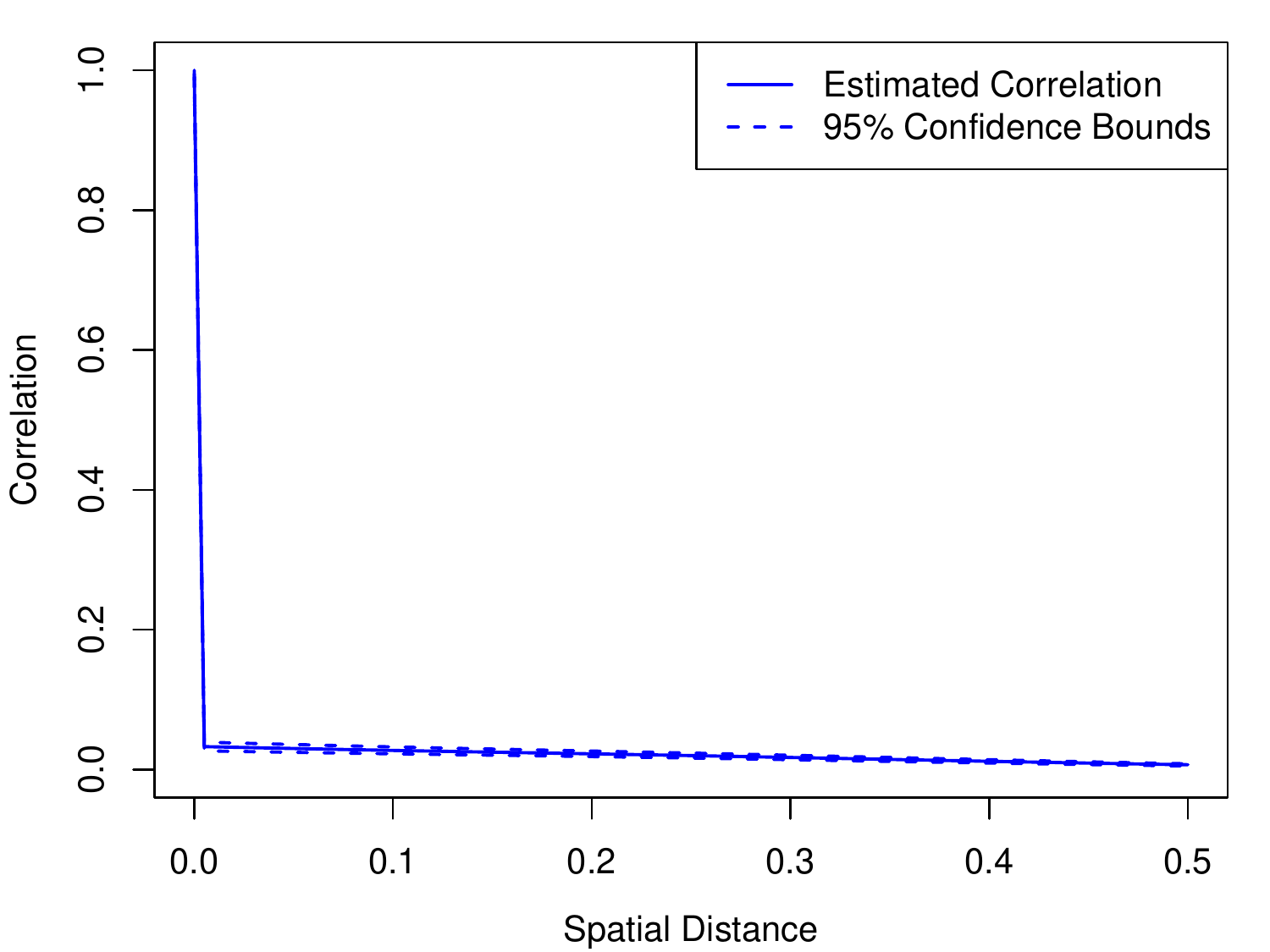}
    \end{subfigure}
    \begin{subfigure}[b]{.49\textwidth}
      \centering
      \vspace{.1in}
      \caption{Tail dependence between nearest spatial \\ \white{.....} neighbors $\delta(s,t)$ and $\delta(s',t)$.}
      \vspace{-.15in}
      \includegraphics[width=.88\textwidth,height=.19\textheight]{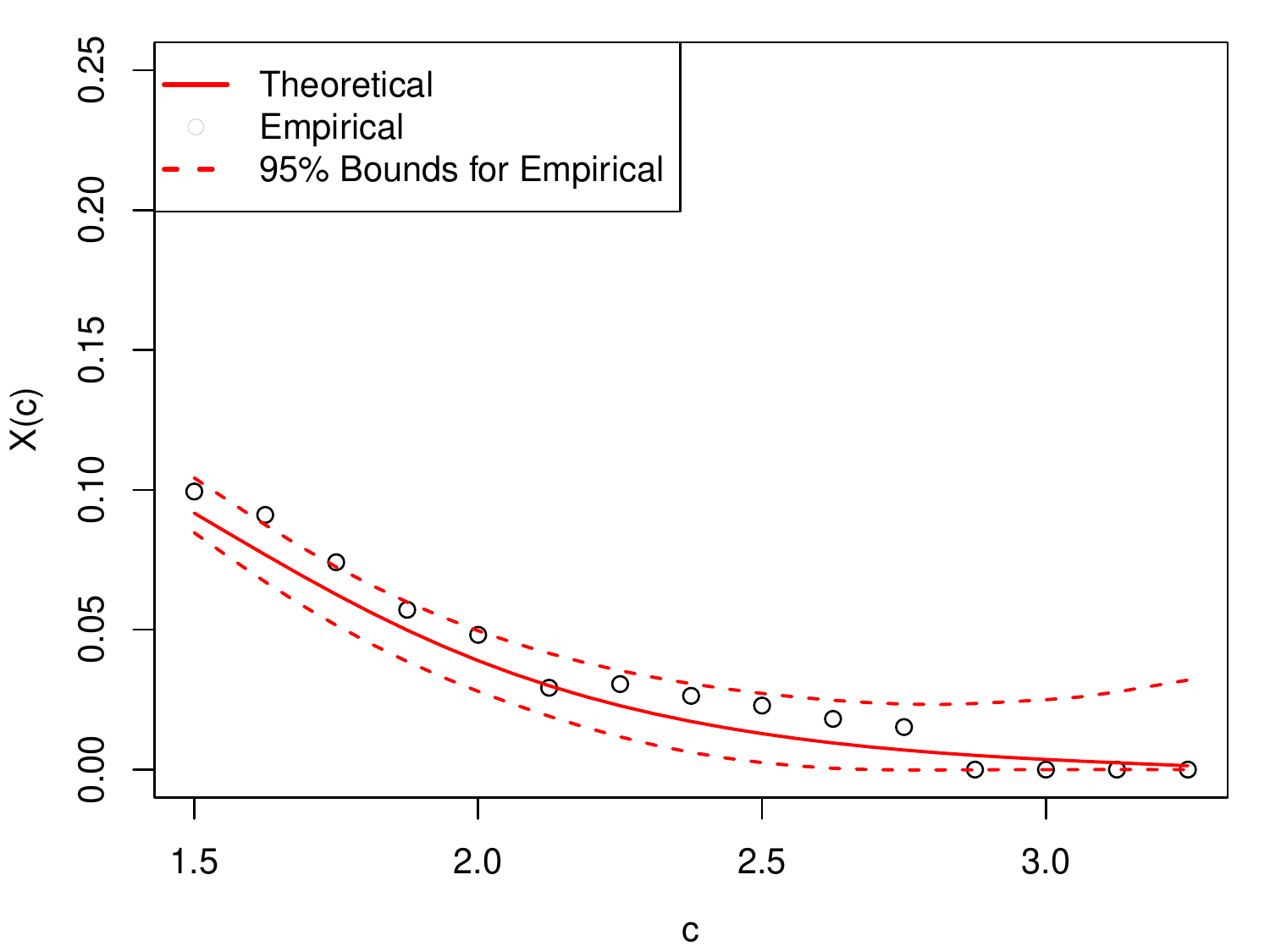}
    \end{subfigure}
\vspace{-.12in}
\end{figure}

The $\delta$ process model appears to allow for the correct tail behavior, marginally, but the properties of the assumed dependency in space and time need to be investigated as well.  Figure~\ref{fig:delta_dependence}(a) provides a plot of the correlation in the underlying GP $\hZ(s,t) = \Phi^{-1}[ F_\delta (\hdelta(s,t))]$ as a function of time lag, estimated empirically from $\hZ(s,t)$ and according to the posterior mean estimate of the exponential model in (\ref{eq:delta_cov}).  While there may be some evidence to suggest the correlation structure of $\hZ(s,\cdot)$ decays slightly differently than exponential, the exponential correlation model is a very reasonable approximation.  As mentioned above the meta-GP does not allow or for dependence in extreme values, i.e., $\chi(c) = \Pr(Z(s,t) > c \mid Z(s,t-1) > c) \rightarrow 0$ as $c \rightarrow \infty$.
However, in Figure~\ref{fig:delta_dependence}(b), the empirically calculated $\chi(c)$ (aggregated over all $s$ and $t$) is plotted across $c$.  The theoretical $\chi(c)$ under a bivariate Gaussian assumption along with 95\% simultaneous bounds (under the Gaussian assumption) on the empirically calculated $\chi(c)$ are also provided for comparison.  The theoretical $\chi(c)$ decays to zero and the empirically calculated values appear to be decreasing as well and are within the confidence bounds.  This implies that the data are not able to inform us of significant tail dependence in this case, i.e., there is not evidence to suggest that a Gaussian copula is insufficient for this analysis.  If this were not the case, a possible extension would be to make use of a $t$ copula as described in \citeasnoun{Demarta05}, for example.

The model in (\ref{eq:node_model}) also assumed no remaining spatial dependence in $\delta$ after accounting for the spatially varying $\beta_j$ coefficients.  Figure~\ref{fig:delta_dependence}(c) displays the correlation in $\hZ(s,t)$ as a function of spatial distance (i.e., a scaled covariaogram).  Distance for this purpose was defined as Euclidean distance from the nodes as they sit in the machine room.  It is clear from Figure~\ref{fig:delta_dependence}(c) that while spatial correlation may be significantly different from zero, it is negligible in the residual process.  Even still, negligible spatial correlation does not necessarily rule out the possibility of asymptotic tail dependence. Figure~\ref{fig:delta_dependence}(d) displays the analog of Figure~\ref{fig:delta_dependence}(b) over space, with $\chi^*(c) = \Pr(Z(s,t) > c \mid Z(s',t) > c)$ estimated empirically, where $s'$ is the closest node to $s$.  There is clearly no evidence of spatial tail dependence and it seems fairly safe to ignore spatial dependence in $\delta$ for this analysis.  Again, if there were non-negligible spatial dependence in $\delta$, a simple extension would be a separable space-time model or one could use any multivariate method such as factor analysis or principle components, etc.

Finally, we conclude this section by comparing the results above to those from an analysis using a simple mixed effects ANOVA, i.e., each $\beta_j(s) \sim N(\mu_j, \upsilon_j^2)$, with continuous AR(1) (i.e., exponential correlation) residuals performed with the \verb1lme1 function from the \verb1nlme1 package in R.  We are not suggesting that this particular analysis would be appropriate here as the residual process has a much heavier right tail than the normal distribution and there is substantial spatial correlation in the regression coefficients $\beta_j(s)$.  This is simply a means to see what may have been gained from the more complex model.  Ignoring these model violations would lead to bias in the $\beta_j$ estimates and the variability involved.  Also, this model still took over an hour to converge for this dataset, so there is not a lot of gain from a computation time perspective.  Some of the estimated $\beta_j(s)$ coefficients from the simple mixed effects model are quite different than those from the proposed model, particularly when there were missing data for some nodes under some conditions.  This was most noteable for the effect due to trays and HPL3.  The spatial dependence should allow for a more reliable estimation in such cases.  Many of the estimates are qualitatively similar, though, and the overall conclusions about the effects due to Wolf, etc., would not be appreciably different.


However, a large concern here is that the goal is to evaluate the state-of-the-machine based on an upper prediction bound for the temperatures that could be achieved.  Thus, the model violation for the upper tail behavior of the residual process is problematic.  The predicted state-of-the-machine using the mixed effects ANOVA model is only 59.8$^\circ\:$C, which is 4.3$^\circ\:$C cooler than that predicted with the proposed model.  This is to be expected in light of Figure~\ref{fig:Norm_GPD_plots}, as a Gaussian assumption for $\delta$ is clearly going to produce overly optimistic (smaller) extreme values.  In any case, this difference is substantial and could easily lead to a poor decision about cooling strategy.  Thus, it is far preferable for this and similar problems to properly account for the tail behavior of the residuals as in (\ref{eq:meta_delta}).


\vspace{-.26in}
\section{Conclusions \& Further Work}
\vspace{-.15in}
\label{sec:conclusions}

An analysis framework for assessing the affect that room changes have on node temperatures in supercomputers has been presented.  This framework was used to assess the effect of cooling system changes and monitor machine room 341 at LANL.  The case study represents a general good-practice for characterizing the effect of cooling changes and monitoring node temperatures in other data centers as layout changes occur.  The framework developed here follows a spatial linear model framework accounting for non-Gaussian heavy tails of the error process via a Gaussian copula and a combination of Gaussian and generalized Pareto for the marginal distribution.  This analysis framework was used to assess the state-of-the-machine after several changes to cooling and uncover the cause of a mysterious overheating episode.  This same framework can also be easily applied to other (larger) data centers as well.  For a given bandwidth of the (permuted) precision matrix, the MCMC algorithm is linear in the number of nodes, and linear in the number of time points per node.  Of course as the number of nodes changes, the layout of nodes has to change and this will likely affect the achieved bandwidth, so in practice the relationship is not that simple.  Still, based on preliminary testing, the analysis of a 20,000 node machine with layout similar to Mustang and 200 time points per node would take $\sim 20$ hours for 20,000 MCMC iterations on a 48 core machine.

The experimental data used were collected during DSTs.  However, it can be difficult to obtain time on a machine during a DST to conduct experiments.  Thus, an alternative approach to gather data is being explored that runs test jobs on nodes when they are not being used by other (regular user) jobs.  The drawback is that it may take some time ($\sim 1$ week) to cycle through all or most nodes in the cluster in this manner.  On the other hand, data could then be collected (at least on part of the machine) nearly continuously.

Very little attention was given in this paper to the question of design of the cooling system for optimization of cost, subject to an acceptable level of node temperatures.  For example, which combination of CRAC units should be used to result in the most efficient cooling of the machines?  What is the optimal cooling supply temperature to use?  The  analysis framework needed to answer these questions, has been developed here, but specific answers to these questions is a subject of future work.

{\small
  \singlespacing
  \vspace{-.23in}
\bibliography{curt_ref.bib}
\bibliographystyle{jasa}
}

\newpage

\begin{appendix}

\setcounter{equation}{0}
\setcounter{page}{1}
\renewcommand{\theequation}{A\arabic{equation}}

\vspace{-.2in}
\begin{Large}
\noindent
{\bf
Supplementary Material: ``Spatiotemporal Modeling of Node Temperatures in Supercomputers''
}
\end{Large}
\vspace{-.3in}

\section{MCMC Algorithm and Full Conditionals}
\vspace{-.05in}
\label{sec:computation}

This section describes the MCMC sampling scheme for the full model described in Section~\ref{sec:model_descr} of the main paper.  The entire collection of parameters to be sampled in the MCMC is
\vspace{-.15in}\beq
\Theta = \left\{
\bbeta,
\bdelta,
\bmu,
\btau,
\blambda,
\upsilon^2,
\theta,
\kappa,
\xi,
\sigma^2
\right\},
\label{eq:param_set}
\vspace{-.15in}\eeq
The MCMC algorithm proceeds with Gibbs updates for each of the elements of $\Theta$ with the exception of $\bdelta$, $\blambda=[\lambda_1,\dots,\lambda_L]'$, and $\theta$, which are updated via a Metropolis Hastings (MH) step.  Full conditional distributions with which to perform the Gibbs updates are provided below for all of the parameter groups listed in (\ref{eq:param_set}) except for $\bdelta$, $\blambda$, $\varphi$, $\upsilon^2$, $\theta$, $\kappa$, and $\xi$, in which case the specifics of the MH step is described instead.\\[-.1in]

\noindent
{$\underline{\bbeta \mid \mbox{rest}}$}

\noindent
Define the collective observation vector for all nodes and all times as $\by=[y(1,1), y(1,2), \dots, y(S,T)]'$ and similarly define the discrepancy vector $\bdelta = [\delta(1,1), \delta(1,2), \dots, \delta(S,T)]'$.  Let the remainder to $\by$ after subtracting off $\delta$ be defined as
\bdm
\bty = \by - \bdelta = \bX \bbeta + \beps
\edm
where $\bbeta=[\beta_0(1), \beta_0(2), \dots, \beta_J(S)]'$.  The design matrix $\bX$ is sparse and can be written as
\bdm
\bX = \left(\bX_0 \mid \bX_1 \mid \cdots \mid \bX_J \right) 
\edm
where each of the $\bX_j$ are block diagonal, e.g., $\bX_0 = \bI_S\otimes \bJ_{\!T}$ with $\bI_{\!S}$ the $S \times S$ identity matrix and $\bJ_{\!T}$ a vector of all ones of length $T$.
Therefore $\bbeta$ conditional on all other parameters and the data reduces to a linear model with variance of the errors known and normal prior on the coefficients. That is, a priori $\bbeta \sim N(\bmu_{\bsbeta}, \bSigma_{\bsbeta})$, where $\bmu_{\bsbeta} = [\mu_0, \mu_1, \dots, \mu_J]' \otimes \bJ_{\!S}$ and
\bdm
\bSigma_{\bsbeta}^{-1} = \bQ_{\bsbeta} = \left(
\begin{array}{cccc}
  \bQ_0 & \bzero & \cdots & \bzero \\
  \bzero & \bR_1 & \cdots & \bzero \\
  \vdots & \vdots & \ddots & \vdots \\
  \bzero & \bzero & \cdots & \bQ_J \\
\end{array}
\right),
\edm
and $\bQ_j$ from (\ref{eq:Q}) is the sparse $S \times S$ precision matrix for $\bbeta_j=[\beta_j(1),\dots,\beta_j(S)]'$ for the current values of $\blambda$ and $\tau_j$. 

The normal distribution is well known to be conjugate in this setting \cite{Gelman2014} and thus
\bdm
\bbeta \mid \mbox{rest} \sim N(\bm, \bV)
\edm 
where
\bdm
\bV^{-1} = \frac{1}{\sigma^2} \bX'\bX + \bQ_{\bsbeta}
\edm
and
\bdm
\bV^{-1} \bm = \left[ \frac{1}{\sigma^2} \bX' \bty + \bQ_{\bsbeta} \bmu_{\bsbeta} \right].
\edm
The resulting precision matrix $\bV^{-1}$ is sparse and efficient means to perform the matrix arithmetic above and generate the multivariate normal for $\bbeta$ exist for such cases (e.g., see \citeasnoun{Rue05} and the \verb1Matrix1 package in R).\\[-.1in]

\noindent
{\underline{MH update for $\bdelta$}}

\noindent
Since there is no spatial dependence, each of the $\bdelta_s$ can be updated independently (and in parallel).  If spatial dependence were required, the same approach as below would work with the product correlation for space/time and the Kronecker structure could be leveraged for computational efficiency.  The MH proposal  $\bdelta_s$ for each $s$ is drawn independently of the current value of $\bdelta_s$ from a conjugate normal update, assuming a priori that $\bdelta_s \sim N(\bzero, \upsilon^2 \bR_\theta^{-1})$, where $\bR_\theta$ is the inverse of the $T_s \times T_s$ time correlation matrix whose $(i,j)$ element is $\exp\{-\theta|t_{s,i}-t_{s,j}|\}$, for observed time points of $t_{s,1}, \dots, t_{s,T_s}$, on node $s$.

Let $\bty_s$ be only the elements of $\bty$ in (\ref{eq:rem_delta}) corresponding to node $s$, and similarly for $\beps_s$ and $\bX_s$.  The remainder to $\by$ after subtracting off $\bX_s \bbeta$ is
\beq
\bty_s = \by_s - \bX_s \bbeta =  \bdelta_s + \beps_s.
\label{eq:rem_delta}
\eeq
Therefore under a normal prior, $\bdelta_s$ conditional on all other parameters and the data would reduce again to a linear model with variance of the errors known and normal prior on the coefficients.  The normal distribution is again conjugate in this setting and thus the proposal for the MH step is provided by
\bdm
\bdelta_s^* \sim N(\bm_s, \bV_s),
\edm
where
\bdm
\bV_s^{-1} = \frac{1}{\sigma^2} \bI_{T} + \frac{1}{\upsilon^2} \bR_\theta,
\edm
and
\bdm
\bV_s^{-1} \bm_s = \frac{1}{\sigma^2} \bty_s 
\edm
Let the (multivariate normal) density of this proposal be denoted $d(\bdelta_s)^*$.  Once again, sparse banded matrix operations make for the efficient evaluation of $d(\bdelta_s)^*$.  The only portion of the full model likelihood that differs between the current value and the proposal is ${\cal L}(\bty_s; \bdelta_s, \sigma^2)$, which is simply {\em iid} normal mean $\bdelta_s$ and variance $\sigma^2$.  Recall the {\em actual} prior for $\bdelta_s$ is the Normal $+$ GPD model marginal with a GP copula, defined in (\ref{eq:meta_delta}).  Let $\pi(\bdelta_s)$ denote the density of this prior distribution which relies on a multivariate normal density evaluation (again with a sparse precision).  The MH ratio for the $\bdelta_s$ update is then
\bdm
MH = \frac{{\cal L}(\bty_s; \bdelta^*_s, \sigma^2)\pi(\bdelta_s^*)d(\bdelta_s)}
{{\cal L}(\bty_s; \bdelta_s, \sigma^2)\pi(\bdelta_s)d(\bdelta_s^*)}.
\edm
\\[-.1in]

\noindent
{$\underline{\mu_j \mid \mbox{rest}, \; j=1,\dots,J}$}
    
\noindent
$\bbeta_j \sim N(\bJ_{\!S} \mu_j, \bQ_j^{-1})$, so that $\bbeta_j = \bJ_{\!S} \mu_j + \bomega$, where $\bomega \sim N(\bzero,\bQ_j^{-1})$. Let $\bQ_j = \bL' \bL$ be the (sparse) Cholesky decomposition.  Then set,
\bdm
\bgamma_j = \bL (\bbeta_j) = \bL \bJ_{\!S} \mu_j + \bL \bomega, 
\edm
where $\bL \omega \stackrel{iid}{\sim} N(0,1)$.  Thus, this is now a simple linear model update of a single regression coefficient $\mu_j$ with known residual variance and design matrix $\bL \bJ_{\!S}$.
A priori, $\mu_j \stackrel{iid}{\sim} N(M_j, S^2_j)$, which is conjugate, leading to the following independent updates for each $j$,
\bdm
\mu_j \mid \mbox{rest} \sim N(m_j, v_j),
\edm
where
\bdm
v_j^{-1} = \frac{1}{S^2_j} + \bJ_s' \bQ_j \bJ_s
\edm
and
\bdm
m_j = v_j \left(\frac{M_j}{S^2_j} + \bJ_s'\bQ \bgamma_j \right)
\edm
\\[-.1in]

\noindent
{$\underline{\tau_j \mid \mbox{rest}, \; j=1,\dots,J}$}

\noindent
$\bbeta_j \sim N(\bJ_{\!S} \mu_j, \bQ_j^{-1})$.  Then set,
\bdm
\bgamma_j = \bL (\bbeta_j - \mu_j) \stackrel{iid}{\sim} N(0,\tau_j^2).
\edm
A priori, $\tau_j \stackrel{iid}{\sim} \Gamma(A_\tau, B_\tau)$, which is conjugate, leading to the following independent updates for each $j$,
\bdm
\tau_j \mid \mbox{rest} \sim \Gamma \left(A_\tau + \frac{S}{2} \;,\;
B_\tau + \frac{1}{2}\sum_{s=1}^S \gamma_{j,s}^2\right).
\edm
\\[-.1in]

\noindent
{\underline{MH update for $\blambda$}}

\noindent
As mentioned in the main paper, $\blambda$ was updated via a MH random walk proposal using a Dirichlet distribution.  The random walk was conducted by drawing a proposal $\blambda^* \sim \mbox{Dirichlet}(\blambda / s)$ for a scale (tuning) parameter $s$. 
The tuning parameter was set to $s=0.015$ to achieve an acceptance rate $\approx$40\%, and resulted in good mixing.  Let the density of the proposal, given the current value of $\blambda$ be denoted $d(\blambda^* \mid \blambda)$.  The only portion of the full model likelihood that differs between the current value and the proposal is $\prod_{j=0}^J{\cal L}(\bbeta_j; \tau_j, \blambda, \varphi)$, the $j\th$ term of which is a multivariate normal density with mean vector $\bJ_{\!S} \mu_j$ and precision matrix $\bQ_j$.  Evaluation of each of these densities is again made efficient by the sparsity of $\bQ_j$.  The MH ratio is then
\bdm
MH = \frac{\left[\prod_{j=0}^J \cL(\bbeta_j; \tau_j, \blambda^*, \varphi) \right]\pi(\blambda^*)d(\blambda \mid \blambda^*)}
{\left[ \prod_{j=0}^J \cL(\bbeta_j; \tau_j, \blambda, \varphi) \right] \pi(\blambda)d(\blambda^* \mid \blambda)},
\edm
where $\pi(\blambda)$ is the density for a Dirichlet$(\ba_\lambda)$ random vector.
\\[-.1in]

\noindent
{\underline{MH update for $\varphi$}}

\noindent
The update for $\varphi$ was conducted via a MH random walk proposal on logit scale.  The random walk was conducted by drawing a proposal $\mbox{logit}(\varphi^*) = \mbox{logit}(\varphi+\epsilon)$ for a deviate $\epsilon \stackrel{iid}{\sim} N(0,s^2)$.
The tuning parameter was set to $s=0.05$ to achieve an acceptance rate $\approx$40\%, and resulted in good mixing.  Let the density of the proposal, given the current value of $\varphi$ be denoted $d(\varphi^* \mid \varphi)$.  The only portion of the full model likelihood that differs between the current value and the proposal is again $\prod_{j=0}^J{\cal L}(\bbeta_j; \tau_j, \blambda, \varphi)$. The MH ratio is then
\bdm
MH = \frac{\left[\prod_{j=0}^J \cL(\bbeta_j; \tau_j, \blambda, \varphi^*) \right]\pi(\varphi^*)d(\varphi \mid \varphi^*)}
{\left[ \prod_{j=0}^J \cL(\bbeta_j; \tau_j, \blambda, \varphi) \right] \pi(\varphi)d(\varphi^* \mid \varphi)},
\edm
where $\pi(\varphi)$ is the density for a Beta$(A_\varphi, B_\varphi)$ random vector.
\\[-.1in]

\noindent
{\underline{MH update for $\upsilon^2$}}

\noindent
The MH update for $\upsilon^2$ is a random walk conducted on the log scale, i.e., $\log({\upsilon^2}^*) = \log(\upsilon^2+\epsilon)$ for a deviate $\epsilon \sim N(0,s^2)$.  A value of $s^2=0.003$ was used to achieve an acceptance rate of $\sim$40\%.  Let the density of the proposal, given the current value of $\upsilon^2$ be denoted $d({\upsilon^2}^* \mid \upsilon^2)$.  The only portion of the full model likelihood that differs between the current value and the proposal is $\prod_{s=1}^S{\cal L}(\bdelta_s; \upsilon^2, \theta, \xi, \kappa)$, i.e., the product of the prior densities for each $\bdelta_s$.  The MH ratio is then
\bdm
MH = \frac{\prod_{s=1}^S{\cal L}(\bdelta_s; {\upsilon^2}^*, \theta, \xi, \kappa) \pi({\upsilon^2}^*)d({\upsilon^2} \mid {\upsilon^2}^*)}
{\prod_{s=1}^S{\cal L}(\bdelta_s; {\upsilon^2}, \theta, \xi, \kappa) \pi({\upsilon^2})d({\upsilon^2}^* \mid {\upsilon^2})},
\edm
where $\pi(\upsilon^2)$ is the density for an IG$(A_\upsilon, B_\upsilon)$ random variable.
\\[-.1in]

\noindent
{\underline{MH update for $\theta$}}

\noindent
The update for $\theta$ follows a completely analogous path as that for $\upsilon^2$.
The MH update for $\theta$ is a random walk conducted on the log scale, i.e., $\log(\theta^*) = \log(\theta+\epsilon)$ for a deviate $\epsilon \sim N(0,s^2)$.  A value of $s^2=0.005$ was used to achieve an acceptance rate of $\sim$40\%.  The MH ratio is
\bdm
MH = \frac{\prod_{s=1}^S{\cal L}(\bdelta_s; {\upsilon^2}, \theta^*, \xi, \kappa) \pi(\theta^*)d(\theta \mid \theta^*)}
{\prod_{s=1}^S{\cal L}(\bdelta_s; {\upsilon^2}, \theta, \xi, \kappa) \pi(\theta)d(\theta^* \mid \theta)}.
\edm
\\[-.1in]

\noindent
{\underline{MH update for $\kappa$}}

\noindent
The update for $\kappa$ also uses random walk conducted on the log scale.  Let the density of the proposal, given the current value of $\kappa$ be denoted $d(\kappa^* \mid \kappa)$.  The MH ratio is then
\bdm
MH = \frac{\prod_{s=1}^S{\cal L}(\bdelta_s; {\upsilon^2}, \theta, \xi, \kappa^*) \pi(\kappa^*)d(\kappa \mid \kappa^*)}
{\prod_{s=1}^S{\cal L}(\bdelta_s; {\upsilon^2}, \theta, \xi, \kappa) \pi(\kappa) d(\kappa^* \mid \kappa)},
\edm
where $\pi(\kappa)$ is the density for an Gamma$(A_\kappa, B_\kappa)$ random variable.
\\[-.1in]

\noindent
{\underline{MH update for $\xi$}}

\noindent
Finally, the MH update for $\xi$ is again a random walk conducted on the log scale.  Let the density of the proposal, given the current value of $\xi$ be denoted $d(\xi^* \mid \xi)$.  The MH ratio is then
\bdm
MH = \frac{\prod_{s=1}^S{\cal L}(\bdelta_s; {\upsilon^2}, \theta, \xi^*, \kappa) \pi(\xi^*)d(\xi \mid \xi^*)}
{\prod_{s=1}^S{\cal L}(\bdelta_s; {\upsilon^2}, \theta, \xi, \kappa) \pi(\xi) d(\xi^* \mid \xi)},
\edm
where $\pi(\xi)$ is the density for an Gamma$(A_\xi, B_\xi)$ random variable.
\\[-.1in]

\noindent
{$\underline{\sigma^2 \mid \mbox{rest}}$}

\noindent
Let $\bE = \by - \bX \bbeta - \bdelta$. Then $\bE \stackrel{iid}{\sim} N(0,\sigma^2)$, and the inverse-Gamma prior on $\sigma^2$ is conjugate, leading to the simple update,
\bdm
\sigma^2 \sim \mbox{IG}\left(A_\sigma + \frac{ST}{2} \;,\;
B_\sigma + \frac{1}{2}\sum_{s=1}^S \sum_{t=1}^{T}E(s,t)^2\right).
\edm\\[-.1in]

\end{appendix}

\end{document}